\documentclass[fleqn,usenatbib]{mnras}

\usepackage{newtxtext,newtxmath}

\usepackage[T1]{fontenc}

\DeclareRobustCommand{\VAN}[3]{#2}
\let\VANthebibliography\thebibliography
\def\thebibliography{\DeclareRobustCommand{\VAN}[3]{##3}\VANthebibliography}

\newcommand{\hi}{\textrm{H\textsc{i}}}
\newcommand{\secref}[1]{\hyperref[#1]{Section~\ref*{#1}}}
\newcommand{\appref}[1]{\hyperref[#1]{Appendix~\ref*{#1}}}
\newcommand{\txteq}[1]{\,{#1}\,}
\newcommand{\Nfg}{N_\text{fg}}
\newcommand{\hMpc}{\,h\,\text{Mpc}^{-1}}
\newcommand{\Mpch}{\,h^{-1}\text{Mpc}}
\newcommand{\degree}{^{\circ}}

\usepackage{graphicx}	%
\usepackage{amsmath}
\usepackage{bm}
\usepackage[dvipsnames]{xcolor}
\usepackage{orcidlink}

\title[Detecting redshift space distortions with MeerKLASS]{Detecting redshift space distortions with forthcoming 21cm intensity maps from MeerKLASS}

\author[D. Tassie et al.]{Daniel Tassie \orcidlink{0009-0000-1440-2578}$^{1}$\thanks{E-mail: daniel.tassie@port.ac.uk},
Steven Cunnington \orcidlink{0000-0001-6594-107X}$^{1}$, 
David Bacon$^{1}$, %
Zhaoting Chen \orcidlink{0000-0002-4965-8239}$^2$,
José Luis Bernal\orcidlink{0000-0002-0961-4653}$^3$
\\
$^{1}$Institute of Cosmology \& Gravitation, University of Portsmouth, Dennis Sciama Building, Burnaby Road, Portsmouth PO1 3FX, UK\\
$^{2}$Institute for Astronomy, The University of Edinburgh, Royal Observatory, Edinburgh EH9 3HJ, UK\\
$^{3}$Instituto de Física de Cantabria (IFCA), CSIC-Univ. de Cantabria, Avda. de los Castros s/n, E-39005 Santander, Spain\\
}

\date{Accepted XXX. Received YYY; in original form ZZZ}

\pubyear{2026}

\begin{document}
\label{firstpage}
\pagerange{\pageref{firstpage}--\pageref{lastpage}}
\maketitle

\begin{abstract}
Redshift space distortions (RSD) provide a powerful measurement of structure formation in modern cosmology. In this work, we explore the potential of 21cm intensity mapping with the MeerKAT Large Area Synoptic Survey (MeerKLASS) to detect RSD and probe gravity within the redshift range of MeerKAT's UHF band $(0.4\txteq{<}z\txteq{<}1.45)$. Using simulated sky observations, we examine the impact from observational effects caused by blind foreground removal and MeerKAT's primary beam, plus $k$-cuts to the measurements. With a multipole expansion formalism, we demonstrate how the quadrupole of the $\hi$ power spectrum can be modelled and also show how a foreground transfer function can reconstruct the distorted signal. With $1,115$ $\rm{deg}^2$ of simulated MeerKLASS observations at $0.4\txteq{<}z\txteq{<}1.0$, our proposed analysis pipeline reveals a $6.7\sigma$ detection of an RSD-induced quadrupole relative to one caused by anisotropic observational effects alone. We also explore parameter inference with the full MeerKLASS $10{,}000$ $\rm{deg}^2$ survey. Within forecasts, we demonstrate how a tomographic analysis across the full UHF redshift range can be leveraged to address certain parameter degeneracies with e.g. the \hi\ bias ($b_\hi$) and cosmic density of $\hi$ ($\Omega_\hi$). By adopting a redshift-dependent model for the bias and \hi\ density combined with simulation-informed priors, we demonstrate how degeneracies can be alleviated, and indicate the potential for MeerKLASS to reach competitive precision on simple probes of modified gravity.

\end{abstract}


\section{Introduction}

Redshift space distortions (RSD) are one of the most powerful measurements of structure formation in modern cosmology. Arising from peculiar velocities along the line-of-sight (LoS), RSDs imprint a characteristic anisotropy in the observed clustering signal that directly encodes the growth rate of cosmic structure $f(z)$. Probing $f(z)$, is one of the key goals of modern cosmology. Constraining this parameter provides a direct way to probe gravity on the largest scales, and measuring $f(z)$ across multiple redshifts allows us test for any deviations from general relativity (GR) \citep{Linder_2013,Baker_2014,Cunnington:2025sdr}, signifying the need for modifications to gravity or the dark sector \citep{Clifton_2012, Joyce_2015}. RSDs provide a direct way to measure $f(z)$ \citep{Huterer_2023}, complementary to distance scale probes such as baryon acoustic oscillations (BAO). Whereas BAO are highly robust to observational systematics \citep{Merz_2021}, RSD measurements are more sensitive to survey specific effects \citep{Samushia_2012} that can introduce anisotropies within the clustering signal.

As we continue to push for greater precision within cosmological surveys, it is becoming more necessary to survey larger volumes in order to suppress cosmic variance. Neutral hydrogen ($\hi$) intensity mapping (IM) has recently emerged as a promising way to map large-scale structures over vast cosmic volumes with high redshift coverage \citep{Bharadwaj_2001, Battye_2004, Wyithe_2007, Chang_2008, Pritchard_2012, Ansari_2012, Santos:20150K, Liu_2020}. Rather than resolving individual galaxies, IM measures the emitted 21cm radiation from the hyperfine `spin-flip' transition of \hi\ \citep{4313902,FURLANETTO2006181,Li:2024fjf}, and while the probability of a $\hi$ atom undergoing this transition is low, this is balanced by the large abundance of cosmic $\hi$. Since we expect most $\hi$ to be contained within galaxies in the post-reionisation Universe \citep{Villaescusa_Navarro_2018}, this proves to be a reliable tracer of the underlying dark matter field. Using radio telescopes to conduct IM of the faint $21$cm signal emitted by \hi\ provides an opportunity to obtain RSD measurements at vastly different wavelengths compared to optical telescopes.

Multiple telescopes that make use of the $\hi$ IM technique are under construction or already in operations. Examples include the Five-Hundred-Meter Aperture Spherical Telescope (FAST; \citealt{Hu:2019okh}), the upgraded Giant Metrewave Radio Telescope (uGMRT; \citealt{Elahi:2024qbp}), the Canadian Hydrogen Intensity Mapping Experiment (CHIME; \citealt{Bandura:2014gwa}), and BAO from Integrated Neutral Gas Observations (BINGO; \citealt{Wuensche_2019}). Many of these are pathfinders for the Square Kilometre Array Observatory (SKAO) \citep{5136190} and MeerKAT \citep{Jonas:2018Jr}, which is the official precursor for the SKA-Mid, has now been in operation for several years. MeerKAT and SKA-Mid can operate in either single dish mode or as an interferometer array. The interferometer array mode operates by sampling the Fourier transform of the primary beam-attenuated sky brightness in the $uv$-plane, which is directly related to the $\hi$ power spectrum. Single dish mode however, measures the specific flux of the 21cm signal with direct imaging, and each dish will act as a single telescope with their maps being co-added \citep{Battye_2013, Bull_2015}. When performing operations in single dish mode, the MeerKAT telescope is a world leading facility for 21cm IM at intermediate redshifts. More specifically, the UHF band of the MeerKAT Large Area Synoptic Survey (MeerKLASS) can probe the $\hi$ density field within the redshift interval $0.4 < z < 1.45$
\citep{MeerKLASS:2017vgf,Cunnington:2025sdr}, with this being extended to a redshift range of $0.35 < z < 3$ when integrated into the full SKAO \citep{2020}. With an array of 64 dishes and a large collecting area, this makes MeerKAT well suited for precision measurements of  the $\hi$ power spectrum.

One of the main challenges within the 21cm IM technique is the problem of foreground contamination, as 21cm IM surveys measure the total radio emission across large areas of the sky over a wide range of frequencies. This means that we collect all contributions to the sky brightness temperature, not only the faint redshifted \hi\ signal, but the much brighter foreground contaminants which are many orders of magnitude brighter than the cosmological \hi\ signal \citep{Alonso:2014dhk, Liu_2020}. The main sources of this foreground emission within the radio range are contributions from galactic synchrotron, free-free emission and extra-galactic point sources. However, the shape of the $\hi$ signal along the frequency direction is highly fluctuated, which in principle allows us to separate it from the foreground emission \citep{Dai_2025, DeCaro:2025qly}. As detailed foreground models remain uncertain, the majority of successful approaches adopt `blind' cleaning strategies. These strategies rely on the spectral smoothness of the foregrounds, rather than prior knowledge of their amplitude and structure \citep{Alonso_2014b,Carucci_2020,Cunnington:2020njn,Spinelli:2021emp}. Principal Component Analysis (PCA) based methods have become one of the most widely adopted foreground cleaning techniques \citep{masui2013measurement,Anderson:2017ert,Cunnington:2022uzo, Carucci:2024qpm,MeerKLASS:2024ypg,Chen:2025bcx}, which exploits statistical decompositions of the data to identify and subtract the dominant smooth modes while preserving the \hi\ signal. PCA based methods, and its extensions such as multi-scale PCA (mPCA), have been validated within simulations \citep{10.1093/mnras/stu792}, and also used in the analysis of MeerKAT intensity maps cross-correlated with optical galaxies from the WiggleZ Dark Energy Survey and GAMA galaxies \citep{Cunnington:2022uzo, Carucci:2024qpm}, as well as the recent detection of the \hi\ auto-power spectrum \citep{MeerKLASS:2026ztn}. However,  due to projecting correlated modes along the LoS, PCA based foreground cleaning methods lead to some cosmological signal loss \citep{Switzer:2015ria, Cheng:2018osq} which can lead to biased recoveries of power spectra if ignored or not correctly accounted for. This signal loss is unavoidable when performing aggressive cleaning but can be corrected for with the use of a transfer function, and has been demonstrated and validated with simulations \citep{Cunnington:2023jpq}. These validations were only performed for the monopole moment of the $\hi$ power spectrum however, thus expanding this to higher order multipoles is a crucial next step. Within this work, we explore this by validating the use of a transfer function on the quadrupole moment of the power spectrum within simulations. We focus on the quadrupole so that we have a direct sensitivity to RSDs and other cosmological information, as in linear theory all information is contained within the first three even multipoles $(\ell =0,2,4)$ and when considering non-linear theory most of the information is predominantly contained within $P_0$ and $P_2$.

Instrumental effects also arise within \hi\ IM, which affects our ability to constrain cosmological parameters. The most notable of which is the effect of the primary beam from MeerKAT, which smooths over density fluctuations transverse to the LoS and suppresses information contained within large $k$-modes (small scales) perpendicular to the LoS \citep{Cunnington:2021czb}. The frequency-dependent primary beam couples to the foreground emission and introduces artificial spectral features that can mimic the 21cm signal \citep{Gkogkou:2025qiq, Matshawule_2021}, or the imprint of the beam can induce its own anisotropic signal similar to one induced by RSDs, which presents an issue for detecting the RSD signal in 21cm IM surveys. In addition to this, beam systematics present a degeneracy between cosmological and nuisance parameters, such as $f(z)$, the linear hydrogen bias $b_{\hi}(z)$, and the modelled beam full width at half maximum (FWHM).

By using mock signal injection to construct a foreground transfer function \citep{Switzer:2015ria,Cunnington:2023jpq,Chen:2025til}, we extend the transfer function reconstruction formalism and explore whether we can accurately recover the quadrupole moment of the $\hi$ power spectrum after blind foreground cleaning, which will provide a pathway of detecting RSD signals with MeerKLASS, which has yet to be achieved with MeerKLASS and other 21cm IM experiments. We then use the reconstructed quadrupole to forecast the constraining power of MeerKLASS under the assumption that the transfer function accurately recovers the anisotropic signal, covering the full redshift range of the UHF-band in a tomographic analysis across six redshift bins. We recast the redshift space \hi\ power spectrum in terms of the RSD parameter $\beta$, and include a beam nuisance parameter $\delta\theta_{\rm beam}$ to capture uncertainties within the beam's FWHM. We focus on degeneracies between $f$ and scale-dependent amplitude parameters such as $b_{\hi}$ and $\Omega_{\hi}$ and the external information required to break them.

This paper is structured as follows: in \secref{sec:modelling} we present the formalism for RSDs, the $\hi$ power spectrum, and instrumental effects that modulate the power spectrum. Within \secref{sec: foreground cleaning and beam} we detail the specifications used to simulate realistic 21cm intensity maps from MeerKLASS as well as the bright radio foregrounds. We then investigate how we can isolate the \hi\ signal from the foregrounds using the PCA cleaning approach and the cosmological signal loss that this inevitably introduces, as well as methods that we can take to reconstruct the lost signal. Following this, \secref{sec: foreground cleaning on quadrupole moment} demonstrates the reconstruction of the quadrupole moment with a foreground transfer function, building upon previous works which have performed this for the monopole moment of the \hi\ power spectrum. We then address the issue of the instrumentally induced anisotropic signal and show how we can achieve a statistically significant detection of a RSD induced signal over this instrumentally induced signal. In \secref{sec:gamma and constraints} we use the reconstructed quadrupole to highlight the potential of MeerKLASS to forecast robust RSD detections in future surveys, and also explore how we can test simple probes of modified gravity, by modelling amplitude parameters as simple polynomials and adopting redshift independent parametrisations of growth in a tomographic approach. We summarise our conclusions in \secref{sec:conclusion}.

\section{Formalism}
\label{sec:modelling}
In the following section we give an outline of the formalism used to model the anisotropic RSD effects on the power spectrum (\secref{sec: RSD}, the $\hi$ power spectrum (\secref{sec:HI PS}), and the observational and instrumental effects (\secref{sec:instrumental effects}).

\subsection{Redshift space distortions}
\label{sec: RSD}
RSDs provide the primary sensitivity to the growth rate of structure in large-scale structure surveys, which makes them a key component in their modelling. RSDs arise from peculiar velocities along the LoS, shifting the apparent positions of dark matter tracers such as galaxies or \hi\ gas. The mapping of the tracer's real space position $(\bm{r})$ and redshift space position $(\bm{s})$ is given by.
\begin{equation}
    \bm{s} = \bm{r} + \frac{\upsilon_z(\bm{r})}{aH(z)}
\hat{z}
\end{equation}
where $\hat{z}$ is the LoS unit vector, $\upsilon_z$ is the LoS velocities, $a$ is the scale factor and $H(z)$ is the Hubble parameter. At large scales, objects tend to fall into high density regions, with the density field becoming squashed. This enhances the clustering amplitude along the LoS, which is the well established Kaiser effect \citep{10.1093/mnras/227.1.1}. The Kaiser term arises from the following assumptions. The first is that in the linear regime $\delta \ll 1$, with $\delta$ being the matter density contrast given by 
\begin{equation}
    \delta(\bm{x}) = \frac{\rho(\bm{x})-\bar{\rho}}{\bar{\rho}},
\end{equation}
so that the peculiar velocity field is irrotational and sourced by the linear density field through the continuity equation
\begin{equation}
    \nabla\cdot \bm{\upsilon} = -aHf\delta.
\end{equation}
The second assumption is the plane-parallel (distant observer) approximation, so that a single LoS direction may be defined for the whole survey volume. The Kaiser effect provides a constant boost to the power spectrum on all scales through the Kaiser term $\left(1 +f(z)\mu^2\right)^2$. Where here $\mu$ is the cosine of the angle down the LoS, which is defined as $\mu = \bm{\hat{k} \cdot \bm{\hat{z}}}$. At smaller scales however, virialised objects have random motions which causes the density field to become stretched, causing the clustering amplitude to become weaker along the LoS. This phenomenon is known as the Fingers of God (FoG) effect \citep{10.1093/mnras/156.1.1P}. The FoG effect is included to model small scale nonlinearities caused by virialised objects. Following the analysis of \cite{10.1093/mnras/sty206}, we adopt the preferential damping term of
\begin{equation}
\label{eq: FoG damping}
    D_{\mathrm{FoG}}(k, \mu, \sigma_{\rm v}) = \frac{1}{1+\frac{1}{2}k^2\mu^2\sigma_{\rm v}^2}  
\end{equation}
where $\sigma_{\rm v}$ is the pairwise velocity dispersion \citep{10.1093/mnras/267.4.1020, 1994ApJ...431..569P, 10.1046/j.1365-8711.2001.04222.x}, which for this work we take to be $400$ km $\mathrm{s}^{-1}$. In \autoref{eq: FoG damping}, $\sigma_{\rm v}$ is purely a phenomenological parameter and is obtained by fitting to the power spectrum and is also assumed to be scale independent which is known to be incorrect \citep{PhysRevD.70.083007, 10.1111/j.1365-2966.2005.09943.x, 10.1093/mnras/stx2971}. However, if $\sigma_{\rm v}$ is weakly scale-dependent over the scales of interest this can be justified \citep{Osinga:2025gpt}. Interestingly, the amplitude of the FoG contribution within \hi\ IM surveys is much different than the FoG contribution in galaxy surveys. As \hi\ gas is only found within relatively massive dark matter halos, this causes the amplitude of the FoG effect to be larger for \hi\ compared to galaxies \citep{Villaescusa_Navarro_2018}.

\subsection{\hi\ power spectrum}

\label{sec:HI PS}

MeerKLASS primarily seeks to detect the $\hi$ power spectrum, whose cosmological signal we can model as, ignoring redshift dependence for brevity,
\begin{equation}
\label{eq: HI ps}
\begin{aligned}
    P_{\hi}(k,\mu) 
    &= \mathcal{A}_{\hi}^2 
    \left( 
        1 
        + \beta\mu^2 
    \right)^{\!2} 
    P_m(k) D_{\mathrm{FoG}}(k, \mu, \sigma_{\rm v})
\end{aligned}
\end{equation}
Where $\mathcal{A}_{\hi}$ is an overall amplitude parameter to represent scale-independent parameters such as the mean temperature brightness of hydrogen gas, $\overline{T}_{\hi}$, and the \hi\ bias $b_{\hi}$. We define this amplitude parameter as $\mathcal{A}_{\hi} \equiv \overline{T}_{\hi}b_{\hi}$. We model $\overline{T}_{\hi}$ by following \cite{Battye_2013}, where
\begin{equation}
\label{eq: hi temp}
    \overline{T}_{\hi}(z) = 180\Omega_{\hi}(z)\frac{h(1+z)^2}{H(z)/H_0}\mathrm{mK}
\end{equation}
where $\Omega_{\hi}$ is the density parameter of $\hi$ with respect to the critical density $(\Omega_{\hi} \equiv \rho_{\hi}/\rho_c)$, $h$ is the dimensionless Hubble parameter $h \equiv H_0/100$, $H(z)$ is the Hubble parameter at a given redshift, and $H_0$ is the present day Hubble parameter. The $\left(1+\beta\mu^2\right)^2$ is included to model contribution from the Kaiser effect with $\beta$ defined as $\beta \equiv f/b_{\hi}$. Finally, $P_m$ is the matter power spectrum obtained with the Boltzmann solver \texttt{CAMB} \citep{Lewis:1999bs}.

In order to detect RSD within IM surveys, we need to detect a statistically significant finding of $\beta\txteq{>}0$, which would indicate a non-zero anisotropic growth rate is present. To achieve a sensitivity to the $\mu$-anisotropies,
we need to expand the model of the \hi\ power spectrum to consider higher-order multipoles. This is because the monopole probes the overall amplitude of the power and better determines the overall bias, whereas the quadrupole probes the anisotropies induced by peculiar velocities which are driven by the gravitational potential and directly links to $f(z)$ \citep{Okumura:2015lvp}. We can consider higher order multipoles by including a multipole expansion within the model of the anisotropic $\hi$ power spectrum, following the analysis outlined in \cite{Cunnington_2020, Bernal_2019, Soares_2021, Berti_2023}. We achieve this expansion through
\begin{equation}
\label{eq: multipole expansion}
    P(k,\mu) = \sum_{\ell} P_{\ell}(k)\mathcal{L}_{\ell}(\mu)
\end{equation}

\begin{equation}
\label{eq: mulitpole expansion int}
    P_{\ell}(k,z) = (2\ell + 1) \int_0^1 \mathrm{d}\mu \, P(k,\mu,z)\mathcal{L}_{\ell}(\mu)
\end{equation}
Where $\mathcal{L}$ are the Legendre polynomials, given by
\begin{equation}
\label{eq: Legendre polynomials}
    \mathcal{L}_0 = 1, \hspace{5pt} \mathcal{L}_2 = \frac{3\mu^2 - 1}{2}, \hspace{5pt} \mathcal{L}_4 = \frac{35\mu^4 - 30\mu^2 + 3}{8},\hspace{5pt} \dots
\end{equation}
which modulate the power within the multipoles.

\subsection{Observational \& instrumental effects}
\label{sec:instrumental effects}

\begin{figure*}
    \centering
    \includegraphics[width=\textwidth]{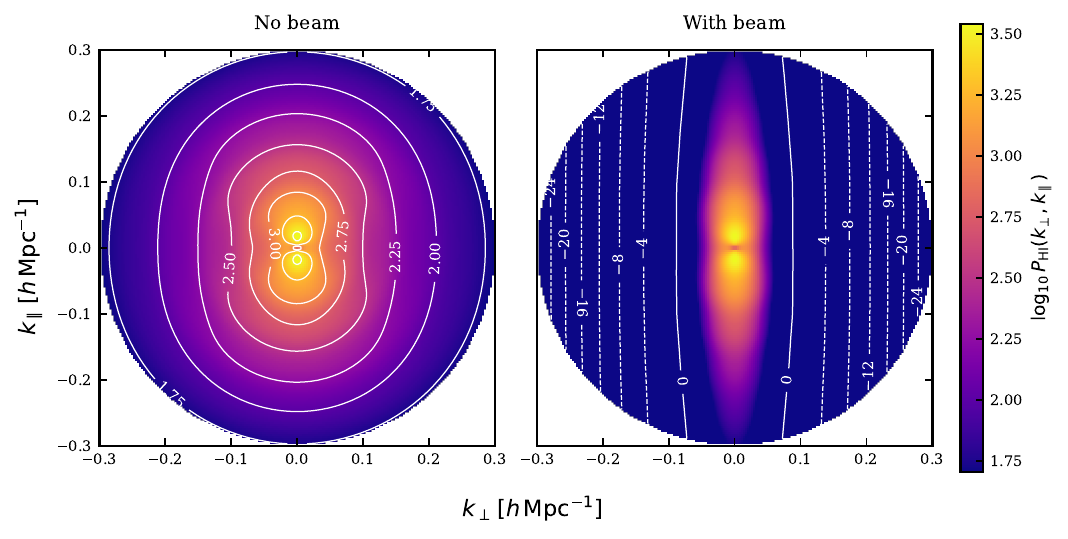}
    \caption{The 2D cylindrical $\hi$ power spectrum at $z=0.75$. The left panel shows the power spectrum without the beam applied to the signal. The RSD effects can be seen in this panel, with the central elongation as a result of the Kaiser effect and the FoG appearing at the outskirts as a squashing effect. The right panel shows the effect of the beam damping the signal. As expected the small scale transverse modes are smoothed over.} 
    \label{fig:figure1}
\end{figure*}

Observational and instrumental effects will also affect the measurement of the $\hi$ power spectrum. Two of the most significant effects will be thermal noise and the telescope's primary beam. We model this in our power spectrum through
\begin{equation}
    P^{\mathrm{obs}}_{\hi}(k,\mu,z) = P_{\hi}(k,\mu,z)\mathcal{B}_{\mathrm{beam}}^2(k,\mu,z) + P_{\mathrm{N}}(z)
\end{equation}
with $P_{\mathrm{N}}$ representing the additive thermal noise, and $\mathcal{B}_{\mathrm{beam}}$ models the primary beam. The beam profile acts as a convolution on the $\hi$ temperature field, which in Fourier space becomes a multiplicative damping of fluctuations in the transverse directions \citep{Battye_2013, 10.1093/mnras/stw3224, Soares_2021, Cunnington_2020, 10.1093/mnras/stac576}. For this work, we assume that the beam profile is Gaussian, with the window given by
\begin{equation}
\mathcal{B}_{\mathrm{beam}}(k,\mu,z)
= \exp\left[-\frac{1}{2} k_\perp^{2} R_b^{2}(z)\right]
= \exp\left[-\frac{k^{2}}{2} (1 - \mu^{2}) R_b^{2}(z)\right]
\end{equation}
where $R_b(z) = \chi(z)\theta_{\mathrm{FWHM}}(z)/2\sqrt{2\ln{2}}$. $\theta_{\rm FWHM}$ is the FWHM profile for MeerKAT which we define by following the steps outlined in \cite{Matshawule_2021}
\begin{equation}
\label{eq: FWHM redshift}
    \theta_{\mathrm{FWHM}} = 1.16 \frac{\lambda(1+z)}{D_{\mathrm{dish}}},
\end{equation}
where $\lambda = 21$ cm and $D_{\mathrm{dish}}$ is the diameter of each dish, which for MeerKAT is $13.5$ m. It has been discussed in \cite{Cunnington:2025sdr} that within a Fisher formalism, the beam term is known exactly and can be perfectly deconvolved from the signal. However, in line with the discussions in \cite{Bernal_2019}, we keep the beam within our signal to more closely match a real data analysis approach.
\autoref{fig:figure1} demonstrates the effects of the beam smoothing on the \hi\ power spectrum. We clearly see the beam creates additional anisotropy, which needs to be disentangled from the anisotropy introduced from RSD.

The noise power $P_{\mathrm{N}}$ is given by the additive thermal noise. We neglect the shot noise which would be minimal for $\hi$ IM \citep{Villaescusa_Navarro_2018, Spinelli_2020}. We therefore model the noise as 
\begin{equation}
    P_{\mathrm{N}} = V_{\mathrm{pix}}\sigma_{\mathrm{N}}^2
\end{equation}
with $\sigma_{\mathrm{N}}$ being the standard deviation of the thermal noise, and $V_{\mathrm{pix}}$ being the voxel volume determined by the pixel area $A_{\mathrm{pix}}$ such that
\begin{equation}
    V_{\mathrm{pix}} = \Omega_{\mathrm{A}}\int_{z_0}^{z_1} \mathrm{d}z\, \frac{\mathrm{d}V}{\mathrm{d}z\mathrm{d}\Omega} = \Omega_{\mathrm{A}}\int_{z_0}^{z_1}\mathrm{d}z \,\frac{c\chi^2(z)}{H(z)}
\end{equation}
where $\Omega_{\mathrm{A}} = A_{\mathrm{pix}}(\pi /180)^2$ and the redshift intervals will be determined by the frequency of observation $\nu$ and the telescope's frequency resolution $\delta\nu$ such that $z_0 = \nu_{\mathrm{21cm}}/(\nu + \delta\nu) - 1$ and $z_1 = \nu_{\mathrm{21cm}}/(\nu - \delta\nu) -1$, with $\nu_{\mathrm{21cm}} = 1420.4$MHz, being the rest-frame frequency of the emitted 21cm radiation from $\hi$.

\section{Simulated 21cm data}
\label{sec: foreground cleaning and beam}
\subsection{Simulation specifications}
Having established a theoretical formalism for the \hi\ power spectrum and the observational and instrumental effects that will modulate it, we now construct realistic mock 21cm intensity maps with log-normal simulations using the \texttt{meer21cm}\footnote{\texttt{meer21cm}: \href{https://github.com/meerklass/meer21cm}{https://github.com/meerklass/meer21cm}} package \citep{2026arXiv260701864C}. Throughout this work, we generate 100 mock lognormal simulations and set up each realisation such that they mimic a UHF-band MeerKLASS survey. To ensure this, each mock realisation spans the redshift range of $0.4<z<1.0$, which corresponds to an approximate frequency range of $714 < \nu < 1020$ MHz\footnote{The full UHF extends to higher redshift ($z\txteq{=}1.45$), but we stick with this reduced subset mainly for computational feasibility. We will forecast the benefit of smaller redshift bins across the full UHF later in the paper}. A frequency resolution of $0.5$ MHz is used, which is consistent with the MeerKAT UHF band receivers. The simulated maps cover a sky area bounded by the right ascension (R.A.) range of $0\degree<\mathrm{R.A.}<40\degree$, and the declinations (Dec.) of $-35\degree<\mathrm{Dec.}<-5\degree$, corresponding to a single patch of sky with an area of $1115\text{deg}^2$, which is representative of the MeerKLASS UHF observations already conducted. The simulations are directly constructed onto the survey light-cone, which allows us to emulate realistic celestial observations. 

MeerKAT's primary beam is then added to our simulations. Each frequency channel is convolved with a frequency-dependent primary beam, whose FWHM profile is described in \autoref{eq: FWHM redshift}. Since the beam size increases with redshift, the physical smoothing scale evolves continuously across the light-cone. We add random Gaussian noise to each simulation realisation to more closely match a realistic scenario. We set this noise at $\sigma_{\rm N}\txteq{=}0.1\,$mK, not dissimilar to current MeerKLASS observations, albeit slightly lower to provide a more robust accuracy test of our pipeline. In principal for higher-order multipoles when assuming the plane-parallel approximation, isotropic noise should just mildly inflate errors and not bias the power.

\subsection{Re-gridding for power spectrum estimation}
\label{sec: beam systematics and foregrounds}
The simulated \hi\ maps represent the observable data products (neglecting foregrounds which we discuss next) from the survey, which are in the celestial coordinate system [R.A., Dec., $\nu$]. Therefore, in order for us to estimate the \hi\ power spectrum, we must re-grid our maps into Cartesian coordinates with dimensions [$l_x, l_y, l_z$] in $\Mpch$. This is an essential step as clustering statistic estimators are most efficiently applied in Fourier space, where Fast Fourier Transforms (FFTs) require data sampled on a regular grid \citep{CooleyTukey}. Recent works have approximated the survey footprint as a rectangular box in comoving space and then applying an FFT directly \citep{Wolz:2015lwa, COMAP:2021sqw}. This approximation introduces some significant biases however, with the power spectrum offset being up to $30\%$ on the largest modes. 

For our simulations, we use \texttt{meer21cm}'s internal regridding routines, which follow the work from \cite{10.1093/mnras/stae333}. This framework incorporates a particle resampling approach that redistributes the voxel intensities of the input sky maps into a 3D Cartesian grid. The \hi\ power spectrum can then be calculated and decomposed into Legendre polynomials according to \autoref{eq: multipole expansion} and \autoref{eq: mulitpole expansion int}. By populating each sky voxel with sampling particles with coordinates that are transformed into comoving space which are assigned to regular grid cells, this approach avoids distortions introduced by large-scale projections. The maps are assigned onto a lower resolution estimation grid, and particles are deposited onto the grid with a mass assignment scheme, which is chosen to be the nearest grid point (nnb for \texttt{meer21cm}), to interpolate the observed \hi\ intensity map.

\subsection{Foreground contamination \& cleaning}
To construct realistic MeerKLASS observations, astrophysical foreground contamination is added to each simulated \hi\ map prior to the power spectrum estimation. For this work, we adopt the blind PCA approach to remove the contaminating astrophysical foregrounds and extract the cosmological signal of interest (see \appref{app: appendix PCA} for more detail). We first begin by generating a mock $\hi$ brightness temperature map, then we generate the foreground emission using \texttt{meer21cm}, which incorporates the \texttt{PySM3} emission map \citep{Thorne_2017, Zonca_2021, Panexp_2025}, which contains contributions \textit{galactic synchrotron} (cosmic rays accelerated by strong magnetic fields within our galaxy), \textit{free-free} emission (free electrons scattering off of ions in our galaxy) and extragalactic \textit{point sources} (e.g. active galactic nuclei). The \hi\ maps are then combined with the foreground maps to mimic what MeerKLASS sees. After these maps are combined together, the primary beam is convolved with the signal. The additive thermal noise is included after beam convolution. The observed sky map is then put through the PCA cleaning process.
\begin{figure*}
    \centering
    \includegraphics[width=\linewidth]{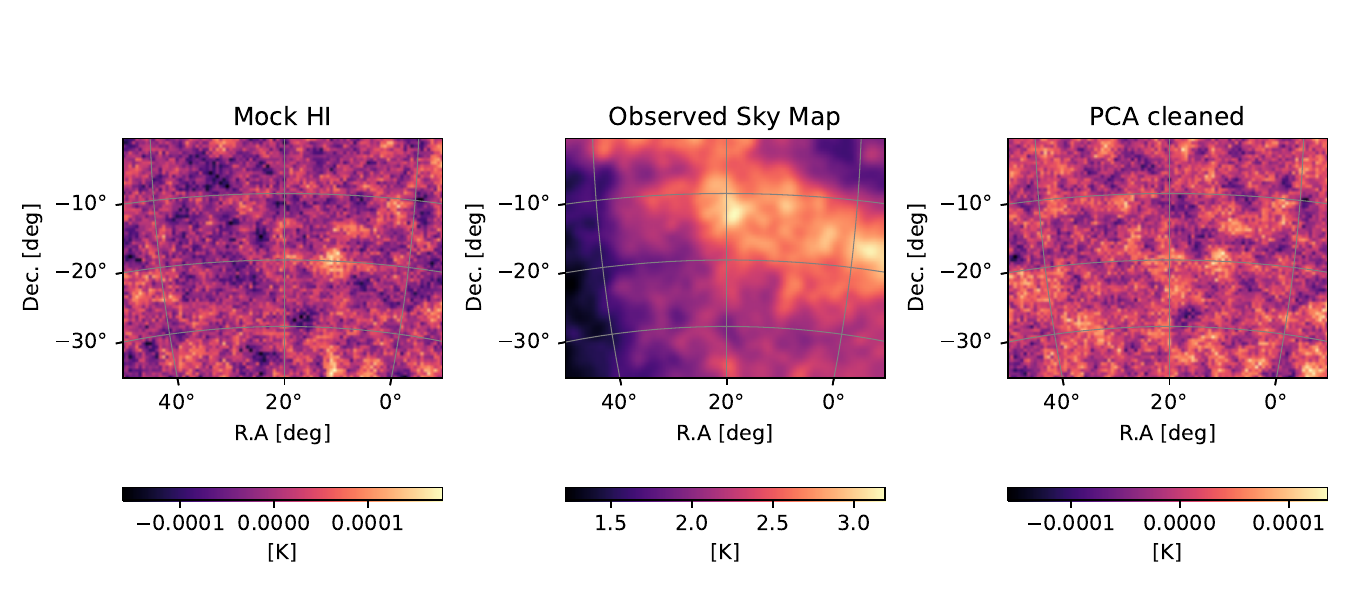}
    \caption{Simulated $\hi$ (\textit{left}), observed sky (\textit{middle}) and residual PCA cleaned (\textit{right}) temperature maps averaged over the first 10 frequency channels, and in the redshift range of $0.4<z<1.0$. We see from the right panel that the PCA cleaning recovers the structure of the original \hi\ signal, however there is some cosmological signal loss, which leads to biased recoveries of power spectra if not correctly accounted for.}
    \label{fig:HI temp map}
\end{figure*}

\autoref{fig:HI temp map} shows the simulated $\hi$, combined foreground and $\hi$ signal, and the PCA cleaned residual temperature brightness, all of which are averaged over the first 10 frequency channels, convolved with the primary beam, and include the additive thermal noise. From the middle panel, which contains the combined contributions from the cosmological signal and the foregrounds, the dominant contribution is the foregrounds, which are around 4 orders of magnitude brighter than the faint $\hi$ signal. The right panel reveals how the PCA cleaning recovers the original \hi\ signal, albeit alongside the signal loss as a result of cleaning, as there is a visible reduction in temperature brightness throughout the simulated map which can bias our measurement of the power spectrum. A key challenge for optimal PCA cleaning is the choice of foreground modes to remove $\Nfg$, which is not fixed a priori. A low choice for $\Nfg$ increases the residual contamination, but a higher value leads to more aggressive cosmological signal loss through over-cleaning. For this work we set our choice to be $\Nfg = 12$. This is likely over-cleaning our mocks, but it should more realistically emulate the aggressive cleaning levels required in current MeerKLASS analyses \citep{MeerKLASS:2024ypg,Chen:2025bcx, MeerKLASS:2026ztn}.

\subsection{The foreground transfer function}

The signal loss from projecting out correlated modes along the LoS is the main consequence of the PCA cleaning approach. This effect is inevitable when aggressively cleaning the maps and must be correctly accounted for to ensure an unbiased power spectrum estimation. Typically, we account for this signal loss through a foreground transfer function $\mathcal{T}(\bm{k})$, which quantifies the attenuation of the true power spectrum due to the cleaning process, defined as
\begin{equation}
    P_{\mathrm{clean}}(k) = \mathcal{T}(k)P_{\mathrm{true}}(k).
\end{equation}
Within previous works \citep{Cunnington:2023jpq}, this has been successfully performed on the monopole moment of the power spectrum. Within this work however, we have a particular focus on reconstructing the quadrupole as the anisotropy from RSDs can be measured here, and the signal-to-noise ratio for higher order multipoles, such as the hexadecapole ($\ell=4)$, will be smaller and carry less statistical significance. The reconstructed quadrupole signal is therefore defined as
\begin{equation}
    P_2^{\mathrm{rec}}(k) = P_2^{\mathrm{clean}}(k)[\mathcal{T}_2(k)]^{-1}.
\end{equation}
$\mathcal{T}_2(k)$ is estimated as the ratio of the auto- and cross-power spectra between the foreground-free and the cleaned mock data, and is constructed following the steps in \cite{Cunnington:2023jpq}. We begin by cleaning the foregrounds from the observed maps,
\begin{equation}
    \textbf{\textsf{X}}_{\mathrm{clean}} = \textbf{\textsf{X}}_{\mathrm{obs}} - \textbf{\textsf{VS}}\textbf{\textsf{V}}^{\textsf{T}}\textbf{\textsf{X}}_{\mathrm{obs}}
\end{equation}
where $\textbf{\textsf{X}}_{\mathrm{obs}}$ are the observed maps, $\textbf{\textsf{V}}$ is the matrix of eigenmodes from the frequency-frequency covariance, and $\textbf{\textsf{S}}$ is a diagonal selection matrix which specifies the $N_{\mathrm{fg}}$ modes to remove. We then generate mock $\hi$ realisations with the same geometry as the data. The mocks are indicated by $\textbf{\textsf{M}}$ to distinguish them from the real data $\textbf{\textsf{X}}$. The mock data is then injected into the \textit{real} data and foreground cleaning is repeated, subtracting the cleaned data without injection to suppress variance, giving the cleaned mock map
\begin{equation}
    \textbf{\textsf{M}}_{\mathrm{clean}} = \left(\textbf{\textsf{X}}_{\mathrm{obs}} + \textbf{\textsf{M}}\right) - \textbf{\textsf{V}}_{\mathrm{obs+m}}\textbf{\textsf{S}}\textbf{\textsf{V}}^{\mathsf{T}}_{\mathrm{obs+m}}\left(\textbf{{\textsf{X}}}_{\mathrm{obs}} +\textbf{\textsf{M}}\right) - \left[\textbf{\textsf{X}}_{\mathrm{clean}}\right].
\end{equation}
The term in the square brackets subtracts the cleaned observed data without the mocks to remove contributions of the map uncorrelated to the mock signal, which therefore reduces variance. The transfer function is then estimated as the ratio of cross- and auto-spectra:
\begin{equation}
    \mathcal{T}_2(k) = \left\langle\frac{\mathcal{P}_2(\textbf{\textsf{M}}_{\mathrm{clean}},\textbf{\textsf{M}})}{\mathcal{P}_2(\textbf{\textsf{M}},\textbf{\textsf{M}})}\right\rangle_{N_{\mathrm{mock}}}.
\end{equation}
Where $\mathcal{P}_2(\bm{X}_a\bm{X}_b)$ represents the operator that measures the quadrupole between two fields (or a single field for an auto-correlation where $a\equiv b$) and bins into band powers, and $N_{\mathrm{mock}}$ is the number of injected mocks used to build the transfer function which for this work we take to be $N_{\mathrm{mock}}=100$ to suppress cosmic variance and ensure convergence of the transfer function. This empirical approach accounts for complex correlations and anti-correlations between signal and foreground modes which would otherwise be neglected, leading to under-estimation of signal loss as demonstrated in \cite{Cunnington:2023jpq}.

\section{Reconstruction Of The Quadrupole}
\label{sec: foreground cleaning on quadrupole moment}
 Foreground cleaning is a necessary step within IM surveys, but it has a significant impact on the anisotropic clustering signal used to probe RSD measurements. We now validate the use of a foreground transfer function to correct for the effects of foreground cleaning on the quadrupole moment of the $\hi$ power spectrum, to analyse how this affects 21cm IM as a probe of RSD. Whilst the formalism that we have outlined is valid for a generic multipole expansion, as outlined before we focus on the quadrupole as higher-order multipoles will carry less statistical significance.

\subsection{Cylindrically averaged power spectrum}
Before presenting the spherically averaged quadrupole, we first present the power spectra in cylindrical-$(k_{\perp}, k_{\parallel})$ space to illustrate the structure of the induced signal loss that we observe, and examine how the transfer function and compensate for this.
\begin{figure*}
    \centering
    \includegraphics[width=\textwidth]{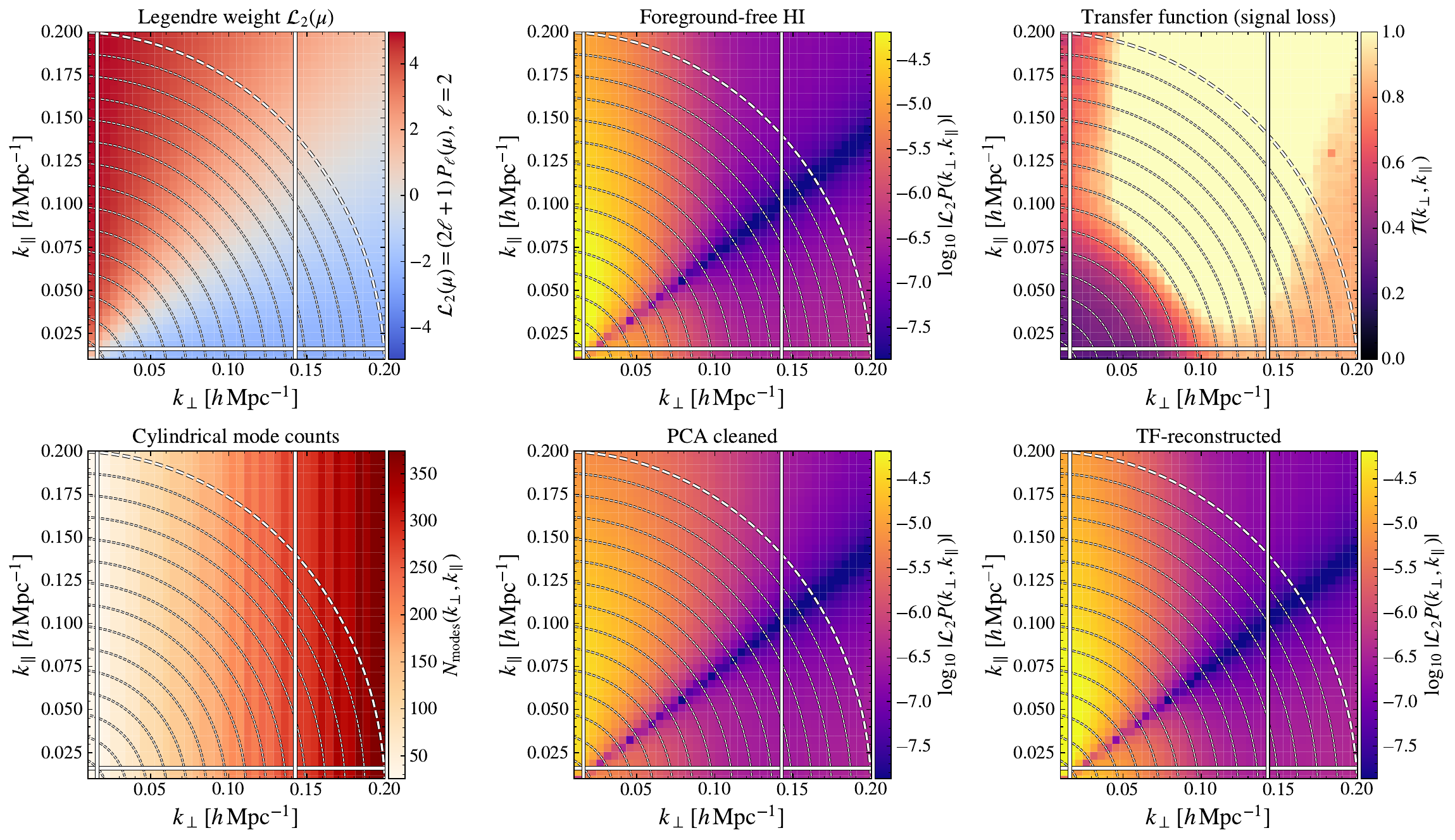}
    \caption{Second order Legendre polynomial $\mathcal{L}_2$ projected onto cylindrical Fourier space (\textit{top left}). Number of modes per cylindrical $k$-bin (\textit{bottom left}). The foreground-free \hi\ quadrupole (\textit{top middle}). PCA cleaned quadrupole (\textit{bottom middle}) demonstrates suppression of large radial modes. The foreground transfer function showing scale-dependent signal loss due to foreground cleaning (\textit{top right}). And the reconstructed quadrupole (\textit{bottom right}) which shows how we are able to recover the lost signal from foreground cleaning. Vertical and horizontal solid white lines represent our chosen cylindrical $k$-cuts discussed in text, and dashed contours show the spherically averaged $k$-bin boundaries}
    \label{fig: six panel plot}
\end{figure*}
The top left panel of \autoref{fig: six panel plot} shows how the second order Legendre polynomial, $\mathcal{L}_2 = (3\mu^2 -1)/2$, which weights the power within the quadrupole, is projected across cylindrical Fourier space, which is crucial for interpreting the structure of the signal loss that we observe. The weighting for $\mathcal{L}_2$ is positive for modes that are aligned with the LoS $(\mu \rightarrow 1)$ and negative for transverse modes $(\mu \rightarrow 0)$. This means that the quadrupole is largely affected by the balance between radial ($k_{\parallel})$ and transverse $(k_{\perp})$ modes, and any anisotropic suppression of modes will significantly influence the quadrupole. 

The bottom-left panel shows the number of cylindrical modes $N_{\mathrm{modes}}(k_{\perp}, k_{\parallel})$. This shows how the sampling of Fourier space is highly anisotropic, with a higher density of $k_\perp$ modes making contributions, thus more weight is effectively given to transverse effects in 1D-averaged multipoles whose $k$-bin boundaries are given the by dashed white contours which we discuss in more detail in the next section, while the largest scale radial modes remain more sparsely populated. 

The impact of the PCA-induced signal loss on the cylindrically averaged quadrupole is shown in the top and  bottom middle panels of \autoref{fig: six panel plot}. These panels demonstrate the signal distortion at the largest scales, and we can clearly see a significant suppression in the signal between the foreground-free \hi\ and PCA cleaned quadrupoles, as the power contained within the largest scales is removed, which leads to a smaller overall amplitude of the quadrupole. Without correction, this would lead to a biased recovery of the quadrupole, and would propagate biases into parameter estimation. We also see that the primary beam and foreground cleaning affect different regions of cylindrical Fourier space. The beam smoothing is most prominent at small transverse scales, while the effects of foreground cleaning are most dominant at large radial scales since the spectrally smooth foregrounds dominate the small $k_{\parallel}$ modes. 

The Legendre weighting explains the structure seen in the transfer function panel in the top right panel of \autoref{fig: six panel plot}. As a consequence, not all Fourier modes will provide equally reliable cosmological information, so to ensure that our analysis is restricted to scales where the signal is well measured and modelling remains robust we apply some scale cuts shown by the solid white lines in \autoref{fig: six panel plot}. We choose our cuts to be $0.016 < k_{\perp} < 0.143 \hMpc$ and $k_{\parallel} > 0.016 \hMpc$. We choose to exclude modes above $k_{\perp}=0.143 \hMpc$ as this scale is determined by the beam's FWHM and scales above this will be dominated by beam smoothing and any information here will be heavily suppressed. Modes contained within $k_{\parallel}<0.016 \hMpc$ and $k_{\perp}<0.016 \hMpc$ will be dominated by cosmic variance, complicated wide angle modes, and small $k_{\parallel}$ regions are where signal loss is most severe and very little information is contained here, hence we also remove these modes. Adding these scale cuts also ensures that we're closely following a realistic data analysis scenario where such scale cuts are required for analyses. While our cuts are motivated by the scale of the beam FWHM, they also closely mimic the ones used in \cite{MeerKLASS:2024ypg} where MeerKLASS maps were cross-correlated with GAMA galaxies.

Analysis of \autoref{fig: six panel plot} suggests that the foreground transfer function is able to reconstruct the true \hi\ quadrupole, recovering the amplitude of the true signal. This indicates that the transfer function is able to successfully recover an unbiased quadrupole signal. Previous work from \cite{Cunnington:2023jpq, Chen:2025til} discuss the use of a transfer function to reconstruct foreground cleaned low redshift IM power spectra. They discuss that even with aggressive foreground cleaning which can lead to a $> 50\%$ bias on the largest scales, the power spectrum can be reconstructed to sub-percent levels of accuracy. Our result is in agreement with the previous work of \cite{Cunnington:2023jpq}, and the level of precision shows strong promise for transfer function reconstruction based approaches for multipole moments of the $\hi$ power spectrum, even aggressive cleaning of our mock data. Given that PCA cleaning introduces anisotropic signal loss through the suppression of small $k_{\parallel}$ modes as demonstrated by the structure of $\mathcal{L}_2$ in \autoref{fig: six panel plot}, the agreement between the reconstructed quadrupole and the true $\hi$ signal indicates the ability of the transfer function to compensate for this $\mu$-dependent distortion.

\subsection{Spherically averaged power spectrum}

\begin{figure}
    \centering
    \includegraphics[width=\linewidth]{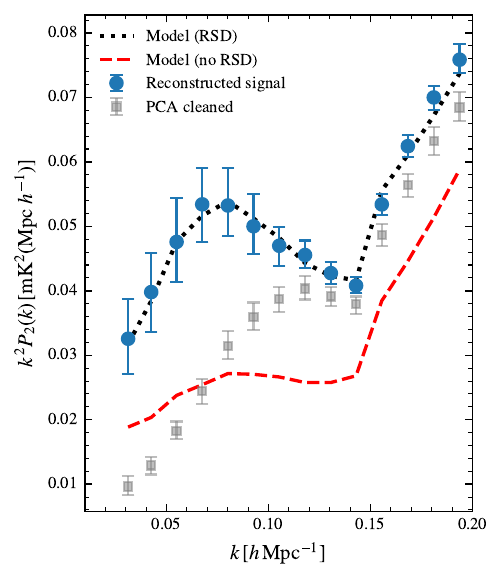}
    \caption{Reconstructed quadrupole moment for the $\hi$ power spectrum. Highlighted in black is the input theoretical model for the \hi\ quadrupole, and in red is the same theoretical model but without any RSD effects ($\beta=0$ and no FoG damping)%
    . In grey is the PCA-cleaned quadrupole, which suffers from severe signal loss at large scales. The blue data points are the foreground transfer function corrected quadrupole averaged over 100 mock realisations, which has been reconstructed to a high degree of accuracy, closely following the input model. We are able to detect the input RSD signature at $6.7\sigma$, which includes effective marginalisation over the power spectra amplitudes.}
    \label{fig:quadrupole reconstruction}
\end{figure}

Having shown that we can reconstruct the anisotropic signal loss in cylindrical Fourier space, we now turn our attention to the spherically averaged 1D quadrupole estimator, which is a more compressed metric and better suited for parameter inference and also allows us to better investigate systematics from the beam. 

\autoref{fig:quadrupole reconstruction} shows the 1D input model for the quadrupole moment before cleaning (\textit{black dotted line}), the PCA cleaned signal averaged over the 100 mock realisations (\textit{grey squares}), and the transfer function reconstructed signal (\textit{blue circles}) with the error bars on the PCA cleaned and reconstructed signal coming from averaging the variance across 100 mock iterations. As discussed before, after performing PCA cleaning there is inevitably some cosmological signal loss as we project out correlated modes. We can clearly see this signal loss within the PCA-cleaned quadrupole at large scales, where a significant suppression relative to the true $\hi$ signal manifests. This is due to the structure of the foregrounds, which are continuum signals and are spectrally smooth through frequency and redshift. As demonstrated in the previous section, this property makes the foregrounds dominate small $k_{\parallel}$ modes, which when removed, also removes some of the cosmological signals on these scales. This leads to the damping of the signal most aggressively on larger scales, particularly large radial scales. 

We also observe an interesting feature in the quadrupole shape, which increases as we push to higher $k$-modes, due to our choice of cylindrical $k$-cuts. As we chose to limit the $k_{\perp}$ modes based on the beam's FWHM, we are limiting the number of transverse modes that are accessible with the selected cut of $k_{\perp} <0.143 \hMpc$. This means we are removing the modes most affected by beam damping, as the beam dominates small transverse scales, leaving radial scales less affected by damping and resulting in a sharp increase in amplitude on these scales. 

\autoref{fig:quadrupole reconstruction} also includes an analytical model without RSD effects (i.e. $\beta = 0$ and no FoG damping) shown in the red dashed line. Ordinarily in a cosmological clustering analyses, we would expect that if there are no RSDs, there would be no quadrupole signal. Interestingly however, there is still a quadrupole signal present, which we mainly attribute to the effects of the primary beam presenting its own anisotropic signal \citep{Cunnington_2020}, but other observational effects, such as re-gridding from celestial to Cartesian coordinates, also contribute to this non-cosmological anisotropic signal. The re-gridding process distorts the anisotropies, since the voxel and cell sizes differ along the LoS compared to angular modes. Our simulations adopt the plane-parallel approximation so $\mu$ is unambiguously defined and the LoS does not vary across the footprint, instead the sampling of $\mu$ is anisotropic. The survey geometry, field-level weights, and the scale cuts applied to the data mean that the modes available in each $k$-bin are not uniformly distributed in $\mu$. This biases the higher-order multipoles more than the monopole, since the monopole is a simple $\mu$-average, whereas the quadrupole isolates the anisotropy through weighted averages over $\mu$ which cancel isotropic contributions via their Legendre polynomial weightings. It is therefore important that we can distinguish between this instrumentally induced signal and the true cosmological signal of interest in order for 21cm IM to probe RSDs. 

\autoref{fig:quadrupole reconstruction} naively suggests an RSD detection is clearly obtainable, given the error bars on the measurement and the difference between RSD and null models. However, since a rescaling of the null model's amplitude could provide a reasonable fit to the data, we must account for the degeneracy with this overall power amplitude to robustly extract the RSD signal. In our case, this means including $b_{\hi}$ and $\overline{T}_{\hi}$ in any fitting, which enter as a scale-independent product that modulates the amplitude of the power spectrum\footnote{We point out that including $\sigma_8$ here as an additional degenerate amplitude parameter would be trivial.}. As mentioned in \secref{sec:modelling}, we use the nuisance parameter $\mathcal{A}_{\hi}\equiv \overline{T}_{\hi}b_{\hi}$ to marginalise over this degeneracy, allowing focus on the anisotropic signal and quadrupole shape that contains the unique RSD contribution. 

Our method for evaluating to what significance we can detect RSD (i.e. $\beta\txteq{>}0$) from the reconstructed quadrupole is as follows. First, $\mathcal{A}_{\hi}$ is fitted for the quadrupole using both with-RSD and no-RSD models. For a given multipole measurement $P_{\ell}(k)$ with uncertainty $\sigma_{\ell}(k)$ we can define
\begin{equation}
    \chi^2 = \sum_{\ell, k} \frac{[P_{\ell}^{\mathrm{data}}(k) - P_{\ell}^{\mathrm{model}}(k;\mathcal{A_{\hi}})]^2}{\sigma_{\ell}(k)^2}
\end{equation} 
where $P_{\ell}^{\mathrm{data}}$ is the multipole measurement, $P_{\ell}^{\mathrm{model}}(k;\mathcal{A}_{\hi})$ is the amplitude-dependent model and $\sigma_{\ell}$ are the error bars. We note that in this $\chi^2$ test we assume that the $k$-bins are not correlated with each other and we direct readers to Figure 6 of \cite{2026arXiv260701864C} for details of the $k$-bin correlation matrices. We fit both the hypothesis (with RSD) model and null (no RSD) model by minimising $\chi^2$ with respect to $\mathcal{A}_{\hi}$.
We then conduct a $\Delta\chi^2$ test for each fitted case, such that
\begin{equation}
    \Delta\chi^2 = \chi^2_{\rm noRSD} - \chi^2_{\rm wRSD}\,,
\end{equation}
which provides the statistical significance of the RSD detection accounting for amplitude degeneracy. This test delivers a detection significance of $\sqrt{\Delta\chi^2}=6.7\sigma$, which indicates that not only can we reconstruct the quadrupole with a foreground transfer function, but we can also do so to a high level of statistical significance over the beam-induced anisotropic signal. 

The importance of this result lies in the fact that the quadrupole has yet to be detected within 21cm IM surveys. Our demonstration is evidence that the foreground transfer function is able to reconstruct the higher-order multipoles of the \hi\ power spectrum, and can be expanded to other multipoles such as the hexadecapole in further work. Furthermore, the quadrupole detection is particularly important as this carries the RSD information. A $6.7\sigma$ detection establishes that this information survives the aggressive foreground cleaning that is required in current MeerKLASS analyses, and is also not suppressed by the primary beam smoothing as well as the anisotropic signal produced by this and other observational effects. We also adopt similar $k$-cuts to that currently used in real MeerKLASS analysis, providing further encouragement that this is feasible with ongoing experiments.

However, whilst we have achieved a highly significant detection, it is worth noting that this is for \textit{mock} data. A scenario with real data where other systematics come into play may reduce this detection significance. On the other hand, a more extensive analysis with real data will provide the opportunity to combine multipoles $(\ell = 0, 2, 4)$, include more redshift bins, and more observation patches which will make any detection of the RSD signal more robust. Overall, our mock data scenario shows promise for detecting RSD signals with MeerKAT's continually growing intensity mapping data.

\section{Potential with full MeerKLASS surveys}
\label{sec:gamma and constraints}
The first MeerKLASS pilot survey was conducted in 2019, with observations in the L-band receiver of MeerKAT \citep{Wang:2020lkn}. Due to the effects of radio frequency interference (RFI) and the fact that large cosmological volumes only become available for redshifted-21 cm at $\lesssim1200$ MHz, the available frequency window was limited to $971<\nu<1023$ MHz corresponding to a useable cosmological redshift range of $0.39<z<0.46$. This motivates the move to the lower-frequency receiver in the UHF-band, which as discussed before offers a dramatically increased redshift window of $0.4<z<1.45$. The move to the UHF-band will provide crucial gains for cosmology, as the observed comoving volume will be $40$ times larger than an equivalent area in the L-band. To date, MeerKLASS has collected over 650 hours of UHF data reaching $\sim3,000$ $\mathrm{deg}^2$. In future surveys, this will increase to $2,500$ hours and a total collected area of $10,000$ $\mathrm{deg}^2$ by 2028 \cite[see][for a summary]{Cunnington:2025sdr}. In the previous section we have demonstrated that we can reconstruct the quadrupole moment of the \hi\ power spectrum with a foreground transfer function, and also detect this reconstructed signal to a $6.7\sigma$ level over a signal induced by observational effects for a single $1,115$ $\mathrm{deg}^2$ observation patch with a thermal noise level of $\sigma_{\rm N} = 0.1$ mK. We now forecast to what extent the full $10,000$ $\mathrm{deg}^2$ MeerKLASS UHF survey will be able to probe RSD when combining the first three higher-order mulitpoles, partioning into finer redshift bins, and considering additional complications from modelling parameters. 

We use the covariance estimated from our lognormal \hi\ simulations and scale it with the number of modes up to the full $10,000\,\rm{deg}^2$ UHF survey to conduct the forecasts. Within these forecasts, we include the first three even higher-order multipoles $\ell=0,2,4$ as most of the cosmological information is predominately contained here. We introduce six tomographic redshift bins (discussed more in \secref{sec:tomography}) spanning the full redshift range of the MeerKLASS UHF band $(0.4<z<1.45)$, and between the scales of $0.02<k<0.3\hMpc$. 

We further explore the effects of scale cuts in our forecasts, comparing an optimistic no-cut test with the realistic cuts chosen for our simulations, as well as the artificial anisotropy that is introduced by the primary beam. When conducting these forecasts, we make the underlying assumption that the signal loss introduced by the PCA cleaning is well contained, and that the signal is recovered by a foreground transfer function without additional bias. This assumption has been validated by our results in \secref{sec: foreground cleaning on quadrupole moment}, where a $6.7\sigma$ detection of an RSD induced quadrupole over a observationally induced signal signifies that the foreground transfer function restores the anisotropic signal. We also explore a number of degeneracies that are introduced within our analyses, in particular we address the degeneracies that arise with $f$ and scale-independent amplitude parameters, which require informative priors on in order to take measurements of $f$. To address these degeneracies, we adopt a tomographic approach, which previous work \cite[e.g.][]{Bernal_2019} has identified as a potential strategy for breaking degeneracies, so we therefore explore this with the full MeerKLASS UHF-band.

\subsection{Parameter degeneracies}
A number of degeneracies will enter into our probes of growth. Here we examine a few of them and explore ways to contain them. Of particular interest is the degeneracy between $f$ and amplitude parameters such as $b_{\hi}$ and $\Omega_{\hi}$. To explore these, we re-introduce the standard parametrisation of the linear redshift space \hi\ power spectrum, given by
\begin{equation}
    P_{\hi}(k,\mu) = \overline{T}_{\hi}^2\left(b_{\hi} + f\mu^2\right)^2P_m(k)\mathcal{B}_{\rm beam}^2(k,\mu),
\end{equation}
Expanding the brackets shows that the signal is dependent on three scale-independent coefficients, each attached to a different power of $\mu$,
\begin{equation}
    P_{\hi}(k,\mu) = \left(\overline{T}_{\hi}^2 b_{\hi}^2
    + 2\overline{T}_{\hi}^2 b_{\hi}f\mu^2
    + \overline{T}_{\hi}^2 f^2{\mu^4}\right)
    P_m(k)\,\mathcal{B}_{\rm beam}^2(k,\mu).
    \label{eq:pkmu_expanded}
\end{equation}
Since that $\overline{T}_{\hi} \propto \Omega_{\hi}$ these coefficients are $(\Omega_{\hi}b_{\hi})^2$, $2(\Omega_{\hi}b_{\hi})(\Omega_{\hi}f)$ and
$(\Omega_{\hi}f)^2$. This means however, that these coefficients are constructed from only two independent quantities being $\Omega_{\hi}b_{\hi}$ and $\Omega_{\hi}f$ and that the three physical quantities $f,\ \Omega_{\hi}, \ b_{\hi}$ cannot be separated by the auto-power spectrum of a single tracer, rescaling $\Omega_{\hi}, \ b_{\hi}$ or $f$ leaves both combinations and therefore the total power spectrum unchanged, which results in a strong degeneracy between $f$ and $b_{\hi}$ as well as $f$ and $\Omega_{\hi}$. This means that the only quantity that we can measure is $\beta = f/b_{\hi}$.
\begin{figure}
    \centering
    \includegraphics[width=\linewidth]{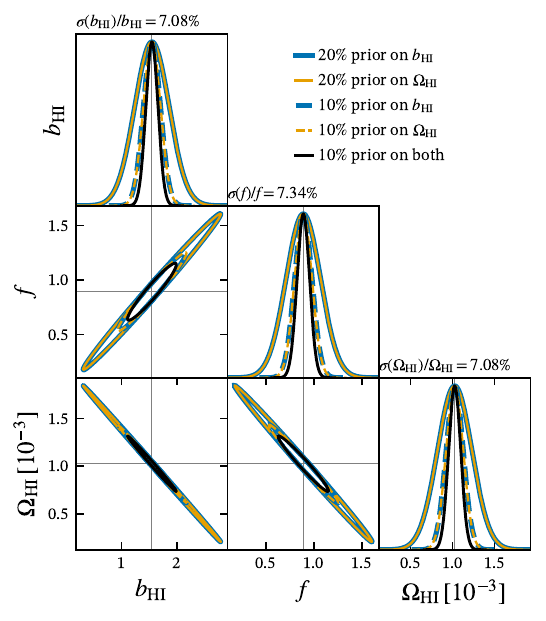}
    \caption{Demonstration of large parameter degeneracies when combining $P_0$, $P_2$ and $P_4$ for a single redshift bin $z=1.1$. The contours correspond to different external Gaussian priors on the amplitude parameters: a $20\%$ prior on $\Omega_{\hi}$ (\textit{dashed orange}) with $b_{\hi}$ free, a $20\%$ prior on $b_{\hi}$ with $\Omega_{\hi}$ free (\textit{solid orange}), a $10\%$ prior on $\Omega_{\hi}$ with $b_{\hi}$ free (\textit{dashed blue}), a $10\%$ prior on $b_{\hi}$ and $\Omega_{\hi}$ free (\textit{solid blue}) and a $10\%$ prior on both amplitude parameters (\textit{black}) with this also being the quoted errors on the parameter panels. This results in cosmological constraints being entirely dependent on prior knowledge of the bias and density of $\hi$, which we seek to address.}
    \label{fig:amplitdue degeneracies}
\end{figure}

\autoref{fig:amplitdue degeneracies} shows the resulting constraints for a single redshift bin forecast at $z=1.1$, from combining the monopole, quadrupole and hexadecapole. The three parameters are almost perfectly degenerate with elongated diagonal ellipses, and therefore require tight priors on them to avoid the ellipses becoming unbound. The marginalised errors $(\sigma(\theta_i)/\theta_i)$ are all pinned by the prior that is imposed on $b_{\hi}$ and $\Omega_{\hi}$, which for this forecast we test a $10\%$ prior on both parameters which gives the quoted error margins in the panel titles in \autoref{fig:amplitdue degeneracies}, $10\%$ on $b_{\hi}$ with $\Omega_{\hi}$ free, a $10\%$ prior on $\Omega_{\hi}$ with $b_{\hi}$ free, a $20\%$ prior on $b_{\hi}$ with $\Omega_{\hi}$ free, and finally a $20\%$ prior on $\Omega_{\hi}$ with $b_{\hi}$ free. As mentioned before, if no priors are imposed then they are perfectly degenerate and the ellipses are unbound, resulting in the marginalised error on $f(z)$ scaling linearly with the prior imposed on the bias or cosmic density of \hi\, meaning that $f(z)$ is only ever as well constrained as $b_{\hi}$ and $\Omega_{\hi}$. 

As outlined in \secref{sec:modelling}, we can address these overall amplitude degeneracies by including the parameters $\mathcal{A}_{\hi}\equiv\overline{T}_{\hi}b_{\hi}$ and $\beta = f/b_{\hi}$, which recasts the form of the \hi\ power spectrum as shown in \autoref{eq: HI ps} albeit without the FoG effect considered. We leave the inclusion of FoG parametrisation and modelling to follow-up work. These have quite unique properties in \hi\ \citep{Villaescusa_Navarro_2018}. Including them in our model would require poorly motivated phenomenological priors on velocity dispersions $\sigma_{\rm v}$ at each redshift, adding further complications and degeneracies. We intend to explore how well bound these terms can be using state-of-the-art hydrodynamical simulations from e.g. \citet{Schaye:2025xuv}.

By adopting the $\beta$ and $\mathcal{A}_{\hi}$ parametrisation, we should also be able to absorb the degeneracy with the matter fluctuation amplitude $\sigma_8$. The nuisance amplitude parameter $\mathcal{A}_{\hi}$ should cover every scale-independent factor. Because the linear matter power spectrum scales as $P_m(k) \propto \sigma_8^2$, the quantity the data actually constrain is the product $\overline{T}_{\hi}b_{\hi}\sigma_8$, the matter fluctuation amplitude $\sigma_8$ is perfectly degenerate with the \hi\ amplitude and is absorbed into $\mathcal{A}_{\hi}$, so this scenario requires no assumption about it. In addition to this, directly constraining $f(z)$ is a difficult task, as it is often constrained as the degenerate product of $f\sigma_8$ from the modelling of galaxy power spectra \citep[e.g.][]{Percival_2009}, and requires combination with another cosmological probe such as the galaxy bispectrum to break degeneracies and tighten constraints \citep{Song_2015, 2015MNRAS.451..539G}. Therefore this parametrisation allows us to conduct forecasts independent of scale-independent parameters by absorbing $\overline{T}_{\hi}b_{\hi}$ into a single nuisance parameter while $\beta$ is measured through the multipole ratios $P_2/P_0, P_4/P_0$. As a result, this allows $\beta$ to be robustly constrained even when the amplitude is left completely free. 

Following this parametrisation and analysing each redshift bin individually whilst marginalising over its amplitude, we forecast a constraint of $\sigma(\beta)/\beta = 6.76\%$ in the lowest redshift bin of $0.40<z<0.575, \ z_{\rm eff} = 0.487$ which corresponds to a detection significance of $\rm{SNR} \approx14.8$, which means that we can directly compare our forecast detections to our simulation result from \secref{sec: foreground cleaning on quadrupole moment}. The forecast therefore presents an improvement in significance by a factor of $\sim 2.2$, reflecting the constraining power of the full MeerKLASS survey volume.

This precision is not maintained across the full band however. The forecast constraint degrades with redshift, from $6.76\%$ in the lowest bin up to $37.9\%$ in the highest bin of $1.275<z<1.45$ which gives a SNR of $\sim\ 2.64$. This corresponds to a degradation of a factor of $5.6$ across the band, which is driven by the combination of a growing beam which damps an increasing fraction of the transverse modes, and the falling signal-to-noise of the $21$ cm signal towards higher redshift with the fixed bin width providing no compensating increase in volume.

Crucially, these are all \textit{single-bin} constraints, obtained by testing each redshift bin as in independent measurement with its own free amplitude. Since the amplitude $\mathcal{A}_{\hi}$ is not infact independent between bins but follows a smooth, physically motivated evolution, this approach discards information. In the following section, we therefore explore what is gained by combining bins into a single tomographic inference, in which the amplitude is described by a shared polynomial across the full redshift range.

\subsection{Tomography}
\label{sec:tomography}

While \hi\ IM introduces additional degeneracies through the mean \hi\ abundance, $\Omega_{\hi}$, it also offers a useful advantage for tomographic modelling. In contrast to optical galaxy surveys, where the observed galaxy bias depends on the details of the target selection, survey depth, colour cuts, completeness, and other instrument-specific effects, intensity mapping measures the integrated 21cm emission from the full unresolved \hi\ population. As a result, the key astrophysical nuisance quantities entering intensity mapping cosmology, namely $b_\hi(z)$ and $\Omega_\hi(z)$, should be slowly evolving, physically consistent functions of redshift that are largely independent of the particular instrument used to observe them. This makes them well suited to joint tomographic modelling with simple smooth parametrisations, which can be progressively informed by increasingly sophisticated simulations, hydrodynamical modelling, and external observational priors. In this section, we therefore motivate a tomographic analysis in which we introduce 6 redshift bins across the full UHF redshift range, and the \hi\ bias and mean brightness temperature are modelled across redshift using simple power-law forms with a small number of nuisance parameters, allowing the redshift evolution of these quantities to be constrained jointly with the cosmological growth signal.

The amplitude, $\mathcal{A}_{\hi} = 180\Omega_{\hi}(z)\frac{h(1+z)^2}{H(z)/H_0}b_{\hi}(z)$, is expected to be a smooth and monotonic function of redshift.
\begin{figure}
    \centering
    \includegraphics[width=\linewidth]{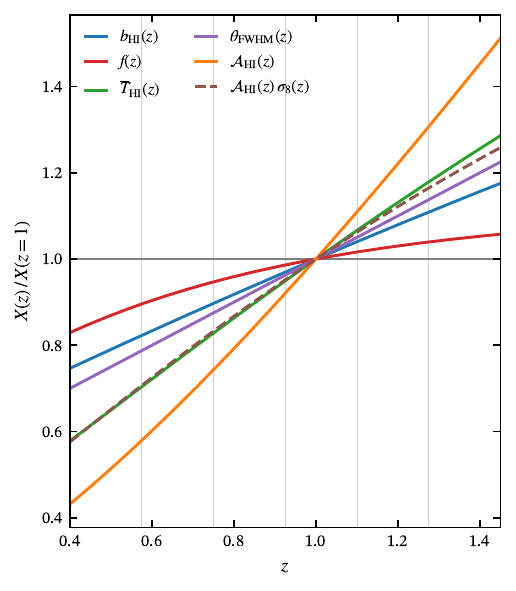}
    \caption{Redshift evolution of the amplitude parameters, $f(z)$ and the beam FWHM, each normalised to their $z=1$ values. Vertical lines represent the tomographic bin edges spanning $0.4<z<1.45$. All four quantities evolve smoothly and monotonically across the full UHF band, differing only mildly in slope so their redshift dependencies are strongly correlated. This similarity limits the leverage of the tomographic approach as neighbouring bins carry little differential information to separate parameters, and means that combining bins can only partially break degeneracies resulting in modest improvements in parameter constraints.}
    \label{fig:Amplitude fitting}
\end{figure}
\begin{figure*}
    \centering
    \includegraphics[width=\linewidth]{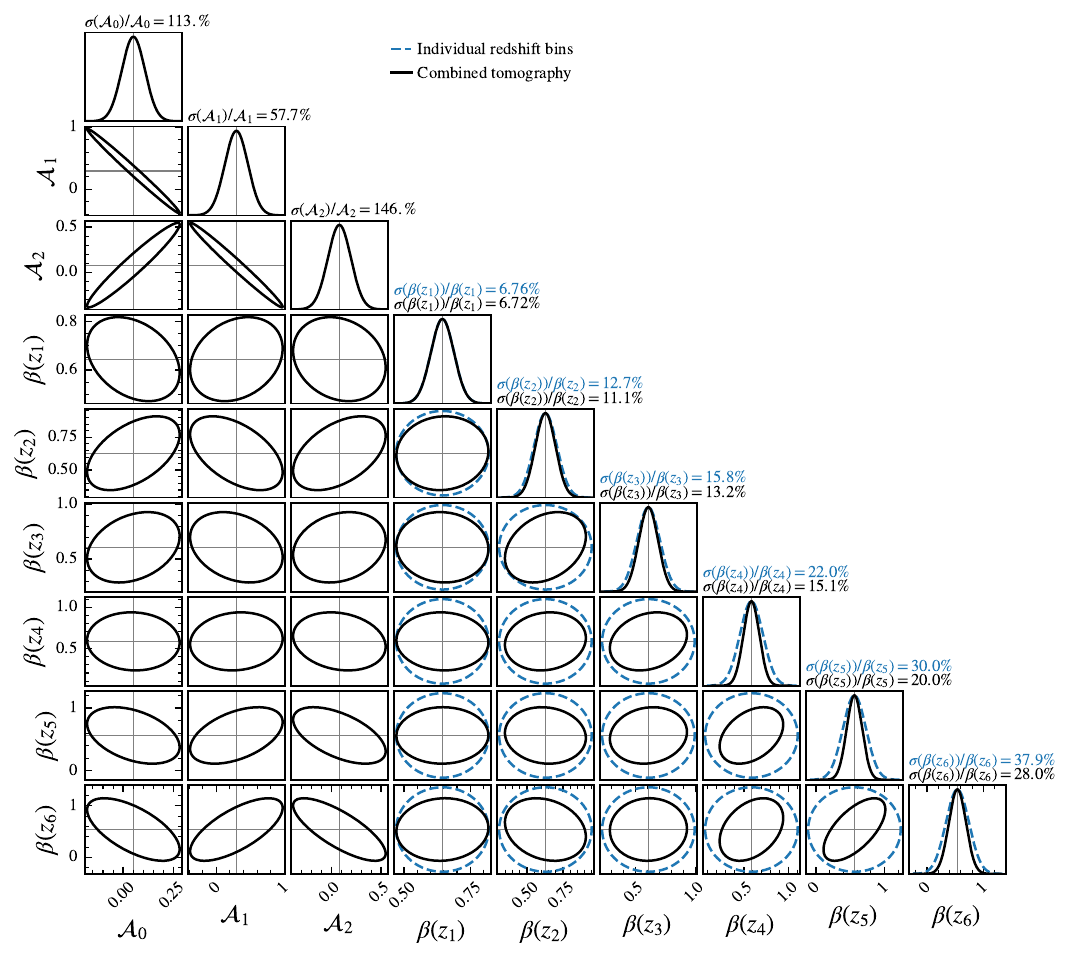}
    \caption{Forecasted constraints from six tomographic bins in the span of $0.4<z<1.45$, with no external priors applied. Dashed blue contours show the constraint from each individual redshift bin where that bin's amplitude is marginalised over as a single free parameter, so at a fixed redshift the polynomial collapses to the single number $\mathcal{A}(z_i)$, therefore describing it with three separate coefficients would be redundant, with the individual bin's amplitude marginalised. Solid black contours show the tomographic constraint where the amplitude coefficients are shared. The amplitude panels are empty for the individual bin case and are constrained by only the tomographic combination. The RSD parameter $\beta$ is recovered to $6.72$-$28.0\%$ across the bins.}
    \label{fig:Tomographic beta forecast}
\end{figure*}
\autoref{fig:Amplitude fitting} shows the evolution of $b_{\hi},\ \overline{T}_{\hi}, \ f,\ \mathcal{A}_{\hi}$ and $\mathcal{A}_{\hi}\sigma_8$ across the full redshift range of the MeerKAT's UHF band each normalised to their $z=1$ value. All four of these parameters are smooth, monotonic and featureless functions of redshift, with each parameter varying in a comparable way across the full redshift range, meaning that their redshift dependences are strongly correlated and do not stand out from one another. The dashed curve shows the combined amplitude parameter $\mathcal{A}_{\hi}$ with the inclusion of $\sigma_8$, which enters as a further multiplicative factor through $P_m(k) \propto\sigma_8$ and is therefore inseprable from the \hi\ amplitude. As $\sigma_8$ drecreases along the redshift band, this flattens the evolution of the grouped amplitude parameter $\mathcal{A}_{\hi}$. We stress however, that $\sigma_8$ is not included as a free parameter in our analyses and is included in \autoref{fig:Amplitude fitting} to demonstrate that our results are independent of any assumption of $\sigma_8$. We also include the evolution of the beam's FWHM, as modelling its effects will also introduce its own uncertainty and in addition to this, the beam's FWHM grows with redshift, meaning that more transverse modes are being smoothed over and decaying the constraining power of the high-redshift bins. The main consequence of this is since the cosmological parameters evolve so similarly, the individual redshift bins carry little differential information with which to separate them, meaning we can only partially break parameter degeneracies. This means that in a tomographic approach we expect to see moderate constraint improvements rather than orders of magnitude improvement. Modelling the amplitude as a quadratic with a small number of coefficients shared across all bins,
\begin{equation}
    \mathcal{A}_{\hi} = \mathcal{A}_0 + \mathcal{A}_1z + \mathcal{A}_2z^2,
\end{equation}
ties all bins together with each bin borrowing amplitude information from one another. Our polynomial fit assumes the combination of the amplitude factors is a smooth function, but early simulations support this fitting and we will further investigate this in future works with higher resolution simulations. This means that each amplitude is no longer free, but constrained to lie on the modelled curve, so each bin has amplitude information from the other bins.

\autoref{fig:Tomographic beta forecast} shows the joint constraints in the following basis $\{\mathcal{A}_1, \mathcal{A}_2, \mathcal{A}_3, \beta(z_1)\dots\beta(z_6), \alpha_{\parallel},\alpha_{\perp}\}$, where $\alpha_{\parallel}$ and $\alpha_{\perp}$ are included to model the Alcock-Paczynski (AP) effect \citep{Alcock:1979mp} and are defined as
\begin{equation}
    \alpha_{\parallel} = \frac{(H(z))_{\rm f}}{H(z)}\qquad \alpha_{\perp}=\frac{D_A(z)}{(D_A(z))_{\rm f}},
\end{equation}
with $D_A$ being the angular diameter distance, and the index $\rm{f}$ representing the fiducial. Each bin contributes its own $\beta_i$ and AP parameters plus the shared amplitude, and the combined matrix is full-rank. The AP effect is included here as within the forecasts we are allowing $\beta=f/b_{\hi}$ to vary, which then means that $\Omega_m$ also varies. Within the methodology for these forecasts, we are looking for phenomenological deviations from a template power spectrum computed in an assumed fiducial cosmology \cite[e.g.][]{Bernal_2019}, that we can then identify with changes in the $\beta$ parameter. On top of $\beta$, the AP effect also introduces anisotropic distortions that are partially degenerate with $\beta$, with this degeneracy changing with redshift. These parameters are related physically through their dependence on $\Omega_m$. Therefore we decide to include the AP effect in our forecast. We explore letting the AP parameters vary freely within the forecasts, and find that by allowing these parameters to vary this significantly degrades the ability to detect $\beta$ within the high-redshift bins and leaves $\beta$ unconstrained in the highest bin with a constraint of $\sigma(\beta)/\beta > 100\%$. This degradation in constraints at high-redshifts happens due to the primary beam removing the transverse shape signature, and $\alpha_{\perp}$ becomes a free per-bin amplitude and marginalising over this direction causes the weakening in constraints. Therefore, in order to adequately measure $\beta$ within forecasts, we opt to place a tight $5\%$ prior on the AP parameters. A $5\%$ prior is looser than what external BAO measurements would deliver on $D_A/r_s$ and $D_H/r_s$, which means that the forecasts do not lean on the prior rather than the actual data. The tomographic sharing allows for information to be pulled from other redshift bins, and can recover the RSD parameter to $\sigma(\beta)/\beta = 6.72$-$28.0\%$ across all bins, while the amplitude coefficients being pure nuisance parameters remain effectively unconstrained at $\sigma(\mathcal{A}_0)/\mathcal{A}_0 = 113\%, \sigma(\mathcal{A}_1)/\mathcal{A}_1=57.7\%,$ and $\sigma(\mathcal{A}_2)/\mathcal{A}_2=146.0\%$. 
\begin{figure}
    \centering
    \includegraphics[width=\linewidth]{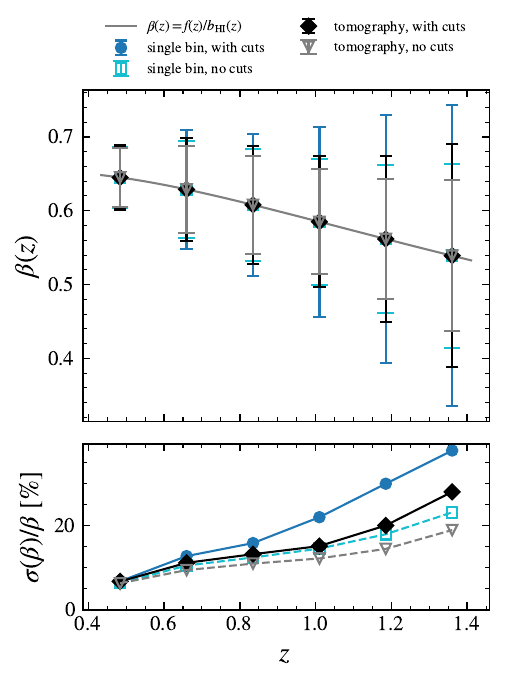}
    \caption{The effect of tomographic combination and of the chosen scale cuts on the recovered RSD parameter constraints. The top panel shows the fiducial model for $\beta$ as the grey curve, and the $1\sigma$ errors in each of the six tomographic bins for a single bin analysis (blue filled circles and cyan open squares) and the full six bin tomographic combination (black filled diamonds and grey open triangles). The filled markers use the scale cuts used in the simulations and represent a more realistic and conservative scenario, whereas the open markers are the optimistic no-cut case. The bottom panel shows the fractional error $\sigma(\beta)/\beta$ for the same cases at each redshift. The tomographic approach tightens the constraints onto $\beta$ in both cases, by a factor that increases with redshift. The scale cuts significantly affect the error bars, with a worsening impact at high-redshifts.}
    \label{fig:beta curve}
\end{figure}

In \autoref{fig:beta curve} we show the forecasted $\beta$ constraints and the marginalised error across the full redshift band of the UHF-survey which isolates two effects. The first is that tying the amplitude across multiple bins tightens $\beta$ relative to an independent bin analysis by a factor that grows with redshift, where individual bin constraints at high-redshifts are the weakest. Secondly, \autoref{fig:beta curve} compares the constraints with and without the simulation scale cuts which are included in the \autoref{fig:Tomographic beta forecast} forecast. The chosen cuts, which remove the small-scale beam-damped transverse modes and the largest scales to contain any signal loss and any wide angle effects, degrade the marginalised error bounds on $\beta$, with the impact worsening with redshift. Applying these realistic scale cuts reduces our tomographic $\beta(z)$ from $4.19$-$8.94\%$ to $6.72$-$28.0\%$ across the six bins. The redshift dependence is driven by the transverse cut. As the beam size grows to larger sizes towards higher redshifts, more of the transverse band is removed, so the highest redshift bins lose more information. The current chosen cuts are conservative, however future analyses with the full MeerKLASS survey is likely to make use of less aggressive cuts in their analysis. We note that we do not include any parametrisation of the uncertainty in the beam FWHM in these forecasts. We omit this parametrisation as these forecasts demonstrate the gains we can achieve from a tomographic approach and lay the foundations for a more realsitic forecast outlined in the next section, where uncertainties in the beam FWHM are taken into account.

\subsection{Probing gravity with the growth rate index}
\begin{figure*}
    \centering
    \includegraphics[width=\linewidth]{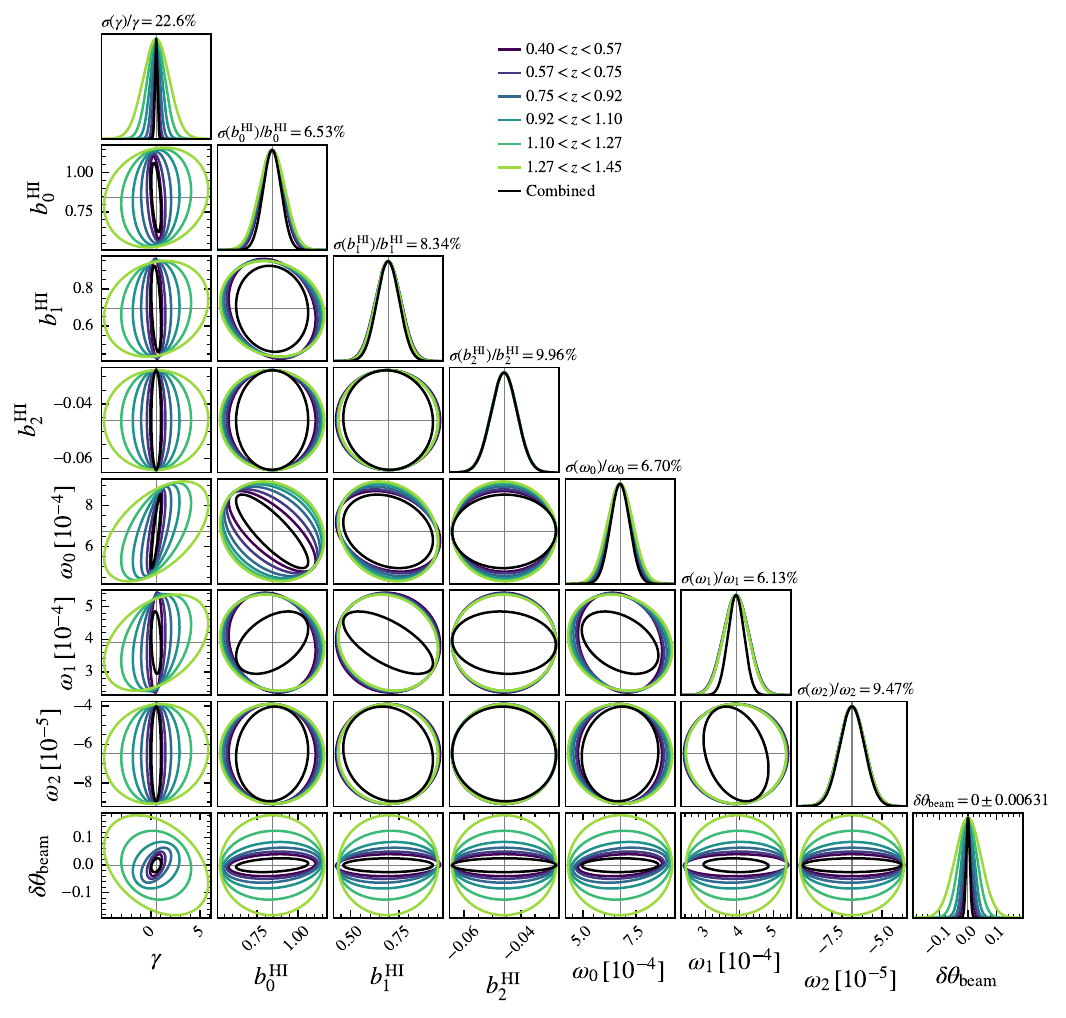}
    \caption{Forecasted constraints on $\gamma$ and the cosmological nuisance parameters from the six tomographic redshift bins in the range of $0.4<z<1.45$. The \hi\ bias and cosmic density are parametrised by the polynomials $b_{\hi}(z) = b^{\hi}_0 + b^{\hi}_1z + b^{\hi}_2z^2$ and $\Omega_{\hi}(z) = \omega_0 + \omega_1 z + \omega_2 z^2$, with the $\Omega_{\hi}$ nuisance parameters shown in units of $10^{-4}$ and $10^{-5}$, and a $10\%$ prior is placed on on all six of the amplitude nuisance parameters. $\delta\theta_{\rm beam}$ is included to parametrise the uncertainty in the FWHM of the beam. Coloured contours represent the individual redshift bin constraints, which constrain parameters weakly with constraints worsening with redshift. The black contours show the combined tomographic constraint, achieving a constraint of $\sigma(\gamma)/\gamma=22.6\%$.}
    \label{fig:gamma forecast}
\end{figure*}
The quantity that we can cleanly measure from the forecasts in \autoref{fig:Tomographic beta forecast} is $\beta = f/b_{\hi}$, however, a more ideal cosmological target is the isolated growth rate $f$. As mentioned before, this is a difficult task due to degeneracies with parameters such as $\sigma_8$ and amplitude parameters such as $\Omega_{\hi}$ and $b_{\hi}$. This therefore means that in order to extract $f$, we need tight external priors on $\Omega_{\hi}$ and $b_{\hi}$ \citep{Obuljen:2017jiy, Castorina:2019zho}. By parametrising $b_{\hi}$ and $\Omega_{\hi}$ as simple power laws that slowly evolve in redshift, we can include them in our tomographic approach. We use the following polynomials to characterise these parameters,
\begin{align}
    b_{\hi}(z) &= b^{\hi}_0+b^{\hi}_1z+b^{\hi}_2z^2\,, \\
    \Omega_{\hi}(z) &= \omega_0 +\omega_1z+\omega_2z^2\,,
\end{align}
where $b^{\hi}_0 \txteq{=} 0.842$, $b^{\hi}_1\txteq{=}0.692$, $b^{\hi}_2\txteq{=}{-}0.046$ and $\omega_0\txteq{=}0.00067$, $\omega_1\txteq{=}0.00039$, $\omega_2\txteq{=}{-}0.000065$. These polynomials as based on hydrodynamical simulations and recent measurements \citep{Villaescusa_Navarro_2018,2020, Cunnington:2022uzo}. To further separate $f$ from $b_{\hi}$ and allow a single cosmological parameter in our tomographic inference, we also switch our parametrisation of growth from $f$ to the growth rate index $\gamma$, which is achieved by using the following approximation \citep{1977ApJ...212L...3S, 10.1093/mnras/251.1.128, PhysRevD.72.043529},
\begin{equation}
    f(z) \approx \Omega_m(z)^{\gamma},
\end{equation}
where $\Omega_m(z)$ is the cosmic matter density given by $\Omega_m(z) = H_0^2\Omega_{m,0}(1+z)^3/H(z)^2$, with $\Omega_{m,0}$ being the cosmic matter density at present day. Switching parametrisation to $\gamma$ provides two advantages. Firstly, as mentioned, it is independent of redshift, so it fits into our tomographic analyses and comes with a convenient combination of information matrices. The second advantage, is that $\gamma$ can act as a simple probe of modified gravity in these forecasts \citep{Fonseca:2019qek, Viljoen:2020efi}, as we have a GR expectation value of $\gamma=0.545$, which therefore means we can investigate the potential of MeerKLASS to place constraints on simple probes of gravity and test if there is room for new physics or modifications to our theories of gravity.

To assess how the $\gamma$ constraint depends on our knowledge of amplitude parameters, we carry out a series of forecasts in which external priors are placed on different subsets of the polynomial coefficients.
\autoref{fig:gamma forecast} shows the achieved parameters constraints for the parameter set of $\{\gamma, b^{\hi}_0,b^{\hi}_1,b^{\hi}_2,\omega_0,\omega_1,\omega_2,\delta\theta_{\rm beam}\}$ with a $10\%$ prior imposed on all six amplitude nuisance parameters $\{b^{\hi}_0,b^{\hi}_1,b^{\hi}_2,\omega_0,\omega_1,\omega_2\}$, which is optimistic but not entirely unrealistic given improving \hi\ simulations and measurements of $\Omega_{\hi}$ \citep{Villaescusa_Navarro_2018, 2020, Cunnington:2022uzo}. We also include a simple instrumental nuisance parameter denoted by $\delta\theta_{\rm beam}$ following the formalism from \cite{Cunnington:2025sdr}, which treats this parameter as a simple perturbation of the FWHM, $\theta_{\rm FWHM} \rightarrow \theta_{\rm FWHM}+\delta\theta_{\rm beam}$. This parameter is then varied alongside the cosmological parameters that are forecasted to mimic the uncertainty in the true FWHM of the beam. We find that modelling the uncertainty in the beam size in our tomographic approach with no external priors, our forecast constraints on $\gamma$ are largely unaffected as the change in the marginalised error on $\gamma$ is $<0.1\%$, indicating that the degeneracy between growth and the modelled size of the beam is minimal. We also chose to omit the AP parameters for these forecasts as we are fixing $\Omega_m$ to a Planck-18 \citep{Planck:2018vyg} value, and we can therefore assume strong prior knowledge of the underlying cosmology. We also find that by fixing the geometry, this does not introduce any bias within our measurements of $\gamma$.

The combined tomography is able to achieve a constraint of $\sigma(\gamma)/\gamma = 22.6\%$. Relaxing the prior unsurprisingly weakens the forecasted constraints, but we find that the priors on the $\Omega_{\hi}$ nuisance parameters are more important than the priors on the bias nuisance parameters. We find that leaving $b^{\hi}_0,b^{\hi}_1, b^{\hi}_2$ completely free within the forecasts yields a constraint onto $\gamma$ of $\sigma(\gamma)/\gamma=29.2\%$, while leaving the $\Omega_{\hi}$ parameters completely free degrades this constraint down to $\sigma(\gamma)/\gamma=33.3\%$, which concludes that constraining the cosmic density of \hi\ is more beneficial than the linear bias, as the redshift-space anisotropy already self-calibrates $b_{\hi}$ more effectively than the amplitude. The headline result from \autoref{fig:gamma forecast} is that the \textit{single} redshift bins achieve weaker constraints on $\gamma$, reaching $\sigma(\gamma)/\gamma=27.9\%$ at $z=0.4$ but being effectively unconstrained at higher redshifts. However, for \textit{combined} redshifts, as polynomial coefficients are shared, the constraint is improved, reaching $22.6\%$. 

We also explore the effects that our chosen $k$-cuts have on our constraints in the same way that we have analysed for $\beta$. The quoted $22.6\%$ is with these conservative cuts applied, however if we remove these cuts and allow the full scale range to propagate into our forecasts we see an improvement on our constraint to $\sigma(\gamma)/\gamma=21.1\%$. The effects of the cuts are less severe here as $\gamma$ is constrained jointly across all six bins through the shared polynomial coefficients, so the degradation of the individual high-redshift bins is partially compensated by the well measured low-redshift end. We consider the forecast containing the $k$-cuts to be our fiducial, since they are physically motivated by the scale of the beam and reflect the scales on which we have reconstructed our quadrupole signal.

\subsection{Outlook and comparison with contemporary probes}

Including degenerate parameters, conservative scale-cuts, an anisotropic beam, and the validation that signal loss from foreground cleaning can be reconstructed, this $\sigma(\gamma)=0.123$ constraint from \autoref{fig:gamma forecast} suggests MeerKLASS holds the potential to distinguish simple braneworld theories of modified gravity such as the Dvali-Gabadadze-Porrati (DGP) model which is often used as a benchmark for the Vainshtein screening mechanism and predicts a value of the growth rate index $\gamma=0.687$ \citep{DVALI2000208,Linder:2007hg, Wei:2008ig}, at a ${\sim}\,1\sigma$ level, and rule out departures larger than $\Delta\gamma = 0.2$ at $2\sigma$. Furthermore, our forecasted precision on $\gamma$ is in line with the results of the Dark Energy Spectroscopic Instrument (DESI) DR1 peculiar velocity survey \citep{Lai:2025xkf}, where they constrain the growth rate of structure from the combination of the bright galaxy survey (BGS) sample, fundamental plane and Tully-Fisher peculiar velocity catalogues combined with the high-redshift growth rate of structure measurements from DESI \textit{ShapeFit} to achieve a constraint of $\gamma=0.580\pm0.110$. This outcome aligns with the expectations from other recent forecasts, which find that next generation surveys will significantly tighten constraints on potential deviations from GR, but will not necessarily detect them unless new physics beyond the $\Lambda$CDM model is present. The J-PAS collaboration \citep{AparicioResco:2019hgh} performed Fisher forecasts on modified gravity parameters, including the effective Newtonian constant $\mu$ and the gravitational slip parameter $\eta$, showing projected sensitivities at the few per-cent level but with central values being consistent with the GR prediction of $\mu = \eta = 1$. Other works such as \cite{Song_2011} explore the complementary of weak lensing and peculiar velocity measurements to test modifications to GR and concluded that error contours on phenomenological modified gravity parameters would shrink significantly around the GR prediction without indicating a strong preference for non-zero deviations in the fiducial case. These studies support the interpretation that future large-scale surveys will be able to perform powerful tests of GR, but could potentially still provide constraints that are consistent with GR unless there is a departure at a level detectable with those surveys.

Forecasts combining weak lensing, cosmic microwave background (CMB) anisotropies and RSDs also further iterate this point. The work from \cite{Ferte:2017bpf} demonstrates that current data already constrain modified gravity parameters to be close to their GR values, and that future survey data will tighten these constraints further without necessarily forecasting a significant deviation if the underlying cosmology follows GR. Comprehensive reviews of cosmological tests of gravity note that whilst future surveys may hold the potential to improve constraints onto phenomenological deviations by orders of magnitude, there remains no anticipated departures from GR. Taken together, our work forecasts that GR remains to be a strong model of gravity on cosmological scales within the sensitivity of current surveys. Our forecasts demonstrate that MeerKLASS holds the potential to place competitive constraints on growth in future surveys, and upcoming data will be able to impose significantly tighter bounds on future experiments. 

We can conclude that within future surveys, leveraging the full $10,000\mathrm{deg}^2$ MeerKLASS survey, we have the potential of more robustly detecting RSD signals and also probing simple probes of modified gravity once cosmological priors are imposed. However, this potential is heavily reliant on external knowledge of $\Omega_{\hi}(z)$ and $b_{\hi}(z)$, as the amplitude degeneracy essentially leaves $f(z)$ unconstrained. Without priors on the polynomial coefficients in the tomographic approach, the constraint onto $\gamma$ significantly weakens, and the strength of the constraint is directly related to the priors imposed on these nuisance parameters, with the $\Omega_{\hi}(z)$ parameters being more impactful. The current priors imposed are motivated from current hydrodynamical simulations but are the limiting factor, which provides a clear path forward for future work with larger volume and higher resolution simulations.

\section{Summary \& Conclusions}
\label{sec:conclusion}
In this work, we have investigated the ability of 21cm IM surveys with MeerKAT to detect RSD signals within higher order multipoles, and explored the forecasting potential of MeerKLASS using the full redshift range of the UHF band $(0.4\txteq{<}z\txteq{<}1.45)$ to forecast constraints on the RSD parameter $\beta$ and explore the potential for future surveys to conduct tests on simple probes of modified gravity. Our analyses have primarily focused on the two main challenges for single-dish IM surveys: the bright contaminating astrophysical foregrounds and MeerKAT's primary beam. We addressed the effects of the bright astrophysical foregrounds, which are many orders of magnitude brighter than the faint cosmological $\hi$ signal, by simulating realistic MeerKLASS maps with the \texttt{meer21cm} python package. We then quantified the effects of the primary beam, and identified a key systematic of the beam presenting its own anisotropic signal. To address the foreground problem, we employed PCA cleaning to separate the contributions from the foregrounds and the \hi\ signal, however using this approach inevitably leads to signal loss as we project out correlated modes along the LoS. This lost signal weakens the amplitude of the anisotropic clustering, which is directly linked to the strength of the RSD signal. To correct for this however, we use a foreground transfer function to reconstruct the signal, which previously had only been validated in simulations for the monopole moment of the $\hi$ power spectrum. Using this approach, we find that not only can we accurately reconstruct the quadrupole signal, but we can do so to a high level of statistical significance over the beam induced quadrupole achieving a detection significance of $6.7\sigma$ over said signal. 

We then explore the potential for MeerKLASS to forecast detections of RSD in future surveys leveraging the full $10,000\mathrm{deg}^2$ UHF survey, as well as testing simple probes of modified gravity within forecasts. Our analyses show that the \hi\ auto-power spectrum only contains the combinations $\Omega_{\hi}b_{\hi}$ and $\Omega_{\hi}f$, so that $f$ carries no independent information and is entirely degenerate with the \hi\ bias and density, and without the presence of prior knowledge on either of these parameters it remains effectively unconstrained. To combat this, we recast the power spectrum in terms of the RSD parameter $\beta = f/b_{\hi}$, which absorbs the the scale-independent amplitude into a single nuisance parameter leaving $\beta$ to be measured cleanly from the shared anisotropy of the signal. Adopting a tomographic approach across six redshift bins and describing the shared amplitude as a quadratic polynomial then allows individual bins to reinforce one another. However, as $\overline{T}_{\hi}(z), b_{\hi}(z), f(z)$ and the primary beam all evolve smoothly and similarly through redshift, the differential information available to the tomographic combination is limited, and the resulting gains are moderate. To connect these measurements to a test of gravity, we parametrise growth through the redshift-independent growth rate index $\gamma$. This provides a robust and efficient way of combining information from multiple redshift bins. By imposing realistic and physically motivated priors on a redshift dependent polynomial for both the \hi\ bias and density, we forecast that MeerKLASS can constrain this simple probe of modified gravity to $\sigma(\gamma)/\gamma=22.6\%$. This result is consistent with measurements from galaxy peculiar velocity surveys from DESI, indicating that MeerKLASS holds the potential to place competitive constraints on growth in future surveys. Crucially, this constraining power is heavily reliant on external priors on $\Omega_{\hi}$ and $b_{\hi}$, and as forthcoming hydrodynamical simulations refine our knowledge of these parameters, the constraining potential for MeerKLASS will strengthen.

Within future works, we can validate our pipeline outlined in this paper by cross-correlating the MeerKLASS $\hi$ power spectrum with optical galaxy surveys to mitigate some of the systematics that we have outlined within this work. One such survey that holds promise would be the luminous red galaxy (LRG) or emission line galaxy (ELG) catalogue from DESI. By including a multi-tracer analysis in future works, this allows us to reduce cosmic variance and significantly boost the signal-to-noise ratio, which is especially powerful for large-scale cosmology, and tests of GR effects. Furthermore, cross-correlations with LRGs or ELGs will provide a more robust detection of the quadrupole, which will improve pur constraints of growth. Looking further ahead, the SKAO will provide a transformative opportunity to extend this work. With an increased collecting area and wider angular surveys, the SKAO will enable 21cm IM surveys to access a significantly larger number of modes over a wider redshift range. This will enhance sensitivity to RSDs and reduce the impact of beam-induced information loss on small scales. 

Overall, this work shows promise for detecting the anisotropies introduced through RSD with 21cm IM with MeerKAT through reconstruction with a foreground transfer function in current and future surveys. We show that we can achieve a statistically significant detection of an RSD-induced quadrupole over the beam-induced signal, and whilst this is for mock data where systematics are not as large as they would be for a realistic case, this detection provides a pathway for a detection of RSD signals within a real data analysis scenario. We further show that adopting a redshift-dependent polynomial form for \hi\ power spectrum amplitude parameters combined, and using $\gamma$ to parametrise growth in a tomographic approach, we can separate the growth rate information from the scale-independent amplitude parameters, albeit requiring moderate prior information on such parameters. Our results provide a clear pathway for detecting RSD signals and show the potential of conducting simple gravity tests with MeerKLASS and future SKAO-era IM surveys, and highlight the importance of jointly modelling cosmological degeneracies and instrumental systematics in future large volume surveys of the large-scale structure of the Universe.

\section*{Acknowledgements}

DT acknowledges support from the Science and Technology Facilities Council (STFC) studentship.
SC acknowledges support from the UKRI Stephen
Hawking Fellowship (grant reference EP/U536751/1). 
ZC is funded by a UKRI Future Leaders Fellowship [grant MR/X005399/1; PI: Alkistis Pourtsidou].
JLB acknowledges funding from the project UC-LIME (PID2022-140670NA-I00), financed by MCIN/AEI/ 10.13039/501100011033/FEDER, UE, and from the project DiffWeb (PID2025-173927NB-I00), financed by MICIU /AEI /10.13039/501100011033.
\section*{Data Availability}

The data underlying this article will be shared on reasonable request to the corresponding author.

\bibliographystyle{mnras}
\bibliography{Bib}

\appendix

\section{Principal Component Analysis}
\label{app: appendix PCA}
PCA cleaning has become one of the most widely adopted foreground cleaning techniques, and has been validated in numerous works and simulations \cite{masui2013measurement, Wolz:2015lwa, Anderson:2017ert, Cunnington:2022uzo, Carucci:2024qpm, MeerKLASS:2024ypg, Chen:2025til}. For each frequency channel $\nu$ we define a map of the total observed brightness temperature $X_{\mathrm{obs}}(\nu, \theta)$ at sky pixel $p$ as
\begin{equation}
    X_{\mathrm{obs}}(\nu,\theta) = T_{\mathrm{FG}}(\nu,\theta) + T_{\hi}(\nu,\theta) + T_{\mathrm{N}}(\nu,\theta)    
\end{equation}
where $T_{\mathrm{FG}}$ are the foreground contaminants, $T_{\hi}$ is the $\hi$ signal and $T_{\mathrm{N}}$ is instrumental noise. The full data cube can then be written in matrix form as
\begin{equation}
    \textbf{\textsf{X}}_{\mathrm{obs}} = \textbf{\textsf{A}}\textbf{\textsf{A}}^{\textsf{T}}\textbf{\textsf{X}}_{\mathrm{obs}} + \textbf{\textsf{R}}
\end{equation}
where $\textbf{\textsf{A}}$ is the $N_{\nu} \times N_{\mathrm{fg}}$ mixing matrix, acting on observations to isolate the contamination which we wish to remove leaving the residual term $\textbf{\textsf{R}}$, containing the cosmological signal and instrumental noise. The cleaning procedure aims to estimate the contaminant contribution $\hat{\textbf{\textsf{A}}}$ and project it out of the data, which gives
\begin{equation}
    \textbf{\textsf{X}}_{\mathrm{clean}} = \textbf{\textsf{X}}_{\mathrm{obs}} -{\hat{\textbf{\textsf{A}}}}{\hat{\textbf{\textsf{A}}}}^{\textsf{T}}\textbf{\textsf{X}}_{\mathrm{obs}}.
\end{equation}
PCA gives an estimate of $\hat{\textbf{\textsf{A}}}$ by identifying a set of eigenvectors $\textbf{\textsf{V}}$ of the frequency-frequency covariance matrix with the largest eigenvalues. These modes capture the majority of the variance in the data, which is assumed to be dominated by spectrally-smooth foregrounds. The first $N_{\mathrm{fg}}$ components are then removed, with the rest being retained as the cleaned data. The choice of $N_{\mathrm{fg}}$ is not fixed as a priori, and is a key challenge for optimal PCA-based cleaning, as a low $N_{\mathrm{fg}}$ increases the residual contamination, but a higher value causes some cosmological signal loss.

\section{Forecast formalism}
\label{sec:forecast formalism}
To forecast the constraints on the parameters considered within \secref{sec:gamma and constraints}, we use the covariance $\textbf{\textsf{C}}$ measured from our lognormal \hi\ simulations and scale this with the number of modes to the full $10,000\,\deg^2$ survey. This provides a representative description of the statistical uncertainties and correlations of the measured multipoles. We therefore perform parameter forecasts by evaluating the the likelihood of the simulated band power data vector under the theoretical model used within the simulations. The forecast matrix is therefore given by
\begin{equation}
    \textbf{\textsf{F}}_{ij}=\frac{1}{2}\mathrm{tr}\left[ \frac{\partial\textbf{\textsf{C}}}{\partial\theta_i}\textbf{\textsf{C}}^{-1}\frac{\partial\textbf{\textsf{C}}}{\partial\theta_j}\textbf{\textsf{C}}^{-1}  \right]
\end{equation}
where $\theta_i,\theta_j$ are from the set of parameters $\Theta$ that influence the derivatives evaluated at their fiducial values $\theta_{\rm fid}$.

For a multipole expansion formalism, the matrix simplifies to \citep{Taruya_2011}
\begin{equation}
    \textbf{\textsf{F}}_{ij}\Big|_{\ell_{\rm max}} = \frac{V_{\rm sur}}{8\pi^2}\sum_{\ell, \ell'}^{\ell_{\rm max}}\int_{k_{\rm min}}^{k_{\rm max}}k^2\, \mathrm{d}k\, \frac{\partial P_{\ell}(k)}{\partial\theta_i}\left[ \tilde{\textbf{\textsf{C}}}^{\ell \ell'}\right]^{-1} \frac{\partial P_{\ell'}(k)}{\partial\theta_j}
\end{equation}
with $\tilde{\textbf{\textsf{C}}^{\ell\ell'}}$ being the reduced covariance.

To compute the derivatives of the forecast matrix, we follow the steps taken in \cite{2020_e} and starting from \autoref{eq: mulitpole expansion int} we have
\begin{equation}
    \frac{\partial P_{\ell}(k)}{\partial\theta_i} = (2\ell +1)\int_0^1 \mathrm{d}\mu \frac{\partial}{\partial\theta_i}[P(k,\mu)\mathcal{L}_{\ell}(\mu)]
\end{equation}
and as the integration limits do not depend on the parameters, we take the derivatives inside the integral. We then use a 5-point stencil to compute the derivatives numerically, defined as a function $\mathcal{F}$
\begin{equation}
    \frac{\partial \mathcal{F}}{\partial \theta} = \frac{-\mathcal{F}(\theta + 2\,\delta\theta) + 8\,\mathcal{F}(\theta + \delta\theta) - 8\,\mathcal{F}(\theta - \delta\theta) + \mathcal{F}(\theta - 2\,\delta\theta)}{12\,\delta\theta}
\end{equation}

\bsp	%
\label{lastpage}
\end{document}